\documentclass[pra,twocolumn,showpacs,preprintnumbers,amsmath,amssymb,floatfix]{revtex4-1}

\usepackage{graphicx}
\usepackage{dcolumn}
\usepackage{bm}
\usepackage{color}

\newcommand{\bra}[1]{\langle #1|}
\newcommand{\ket}[1]{|#1\rangle}

\newcommand{\affA}{Optics and Photonics Research Center, Department of Physics, Institute for Advanced Studies in Basic Sciences (IASBS),Gava Zang, Zanjan 45137-66731, Iran}
\newcommand{\affB}{Zentrum f\"ur Optische Quantentechnologien, Fachbereich Physik, and Hamburg Center for Ultrafast Imaging, Universit\"at Hamburg, Luruper Chaussee 149, 22761 Hamburg, Germany}

\begin{document}



\title{
Compound atom-ion Josephson junction: \\Effects of finite temperature and ion motion
}


\author{Mostafa R. Ebgha}\email[E-mail: ]{mrebgha@iasbs.ac.ir}
\author{Shahpoor Saeidian}
\affiliation{\affA}
\author{Peter Schmelcher}
\author{Antonio Negretti}
\affiliation{\affB}

\date{\today}


\begin{abstract}
We consider a degenerate Bose gas confined in a double-well potential in interaction with a trapped ion in one dimension and investigate the impact of two relevant sources of imperfections in experiments on the system dynamics: ion motion and thermal excitations of the bosonic ensemble. Particularly, their influence on the entanglement generation between the spin state of the moving ion and the atomic ensemble is analyzed. We find that the detrimental effects of the ion motion on the entanglement protocol can be mitigated by properly choosing the double-well parameters as well as timings of the protocol. Furthermore, thermal excitations of the bosons affect significantly the system's tunneling and self-trapping dynamics at moderate temperatures; i.e., thermal occupation of a few double-well quanta reduces the protocol performance by about 10\%. Hence, we conclude that finite temperature is the main source of decoherence in such junctions and we demonstrate the possibility to entangle the condensate motion with the ion vibrational state. 
\end{abstract}

\maketitle


\section{Introduction}
\label{sec:intro} 

The recently attained experimental controllability for generating and manipulating atomic quantum mixtures affords a new playground to study many-body quantum physics. Some examples are: the formation and spread of correlations, mediated interactions and polarons~\cite{KoschorreckNature12,FukuharaNP13,Grusdt2014,MassignanRPP2014,JorgensenPRL16,HuPRL16,CetinaScience16,MeinertScience17,ScazzaPRL17,Rubio2018,GrusdtPRX18,CamachoPRX18,SchmidtRPP2018}. Of particular interest is the generation of genuine quantum correlations that are of relevance for applications such as metrology~\cite{PezzeRMP18} and sensing~\cite{DegenRMP17} as well as for fundamental research related, e.g., to the classical-to-quantum transition~\cite{ZurekRMP03}. Within the plethora of compound atomic quantum systems recently realized in the laboratory (Bose-Fermi mixtures, spinor condensates, etc.), atom-ion systems constitute a rather unique platform because of the long-ranged interspecies interaction -- on the order of a few hundred nanometers -- compared to ultracold neutral matter, where the interaction range is on the order of a few nanometers. The competing effects owing to the different length and energy scales involved in the system enable one to study, as a paradigmatic example, the polaron strong-coupling regime more naturally~\cite{CasteelsJLTP11}. They also give rise to phenomena like density bubbles~\cite{MassignanPRA05,GooldPRA10} and the formation of mesoscopic molecular ions~\cite{CotePRL02,SchurerPRL17}, to mention only a few (we refer to Refs.~\cite{HarterCP14,WillitschFermi15,CoteAAMOP16,TomzaRMP19} for comprehensive overviews on the subject). In all these instances, the crucial role played by correlations beyond mean-field theory is the primary feature of such multilength and multienergy scale physics. 

Among the cornucopia of systems exhibiting macroscopic quantum behavior, atomic Josephson junctions (JJ) have acquired special attention in the last few years and offer unique prospects to study out-of-equilibrium dynamics in a very controllable manner. For example, measurements of high-order correlation functions~\cite{SchweiglerNature17} and quantum transport~\cite{ValtolinaScience15,BurchiantiPRL18,PigneurPRL18} can be carried out. In the former case, the system turns out to be a special-purpose quantum simulator of the so-called sine-Gordon model, which is a relevant integrable model for interacting quantum field theories, that enables access to nonperturbative information about different interesting quantities such as correlation functions or excitation spectra~\cite{NiccoliJST10}. 
Furthermore, it has been theoretically shown that the tunneling dynamics in bosonic JJs (BJJs) can be controlled with an impurity like a trapped ion enabling, e.g., the engineering of tailored entangled states between the ionic internal degrees of freedom and the motional states of a Bose-Einstein condensate (BEC)~\cite{GerritsmaPRL12,SchurerPRA16}. Particularly, it has been shown that by accurately choosing the atom-ion scattering length it is possible to induce macroscopic self-trapping, namely, suppression of tunneling through the barrier. Such a phenomenon was predicted to occur only by controlling the relative phase between the condensates in the two wells and the interspecies interactions~\cite{SmerziPRL97}. Hence, such a capability would allow for the generation of large many-particle impurity states useful, e.g., for inferring scattering properties of the compound system via interferometric measurements as well as for information processing tasks~\cite{DoerkPRA10,SeckerPRA16}. Moreover, such ion-controlled BJJs can be viewed as building blocks of quantum simulators of condensed-matter~\cite{BissbortPRL13,NegrettiPRB14,Gonzalez-CuadraPRL18} and lattice gauge models~\cite{DehkharghaniPRA17}.  

In previous studies of such controlled junctions, the following systems were investigated in detail: For the case of a single atom and single ion in the presence of micromotion and imperfect ion ground-state cooling~\cite{JogerPRA14}, it was found that a large ion-atom mass ratio and a minimal atom-ion separation considerably reduce the detrimental effects of the ion micromotion, similar to what has been pointed out in Refs.~\cite{CetinaPRL12,NguyenPRA12,KrychPRA15}. Many bosons interacting with a static ion in the framework of the two-mode Bose-Hubbard~\cite{GerritsmaPRL12} and with numerical \textit{ab initio} simulations~\cite{SchurerPRA16} have also been investigated. In particular, Ref.~\cite{SchurerPRA16} has confirmed that within the execution time of the ion-BEC entanglement protocol, mean-field theory describes very accurately the self-trapping dynamics, whereas the two-mode approximation~\cite{MilburnPRA97} with time-dependent orbitals describes the tunneling regime very precisely. Specifically in the tunneling regime, a natural population analysis (i.e., computation of the eigenvalues $\lambda_j$ of the one-body density matrix) shows that a second orbital becomes populated up to 5\% (i.e., $\lambda_2\le 0.05$), while at most 95\% of the atoms are in the condensate mode (i.e., $\lambda_1\ge 0.95$). Hence, the occupancy of a second orbital is indeed small, that is, mean-field and Bogoliubov theory are quite good descriptions of the system dynamics. Therefore, up until now, the system was investigated only (1) when the ion was tightly trapped so that its motion could be neglected and (2) at zero temperature. While case 1 could be justified to some extent if the ion trap frequency is much larger than the atom trap frequency, case 2 is much harder to attain, as in experiments thermal fluctuations are unavoidable. Hence, the present paper aims to include these effects in the description of the ion-controlled BJJ. Towards that end, we derive equations of motion for the compound atom-ion systems by starting from a many-particle wave-function ansatz that is simpler than that of \textit{ab initio} methods at zero temperature~\cite{OfirPRA08,CaoJCP13,KroenkeNJP13}, where all correlations are taken into account. More precisely, in our ansatz we shall assume that, regardless of the ion motional state, the bosonic quantum state can be described by a product state that relies upon the occupied eigenstate of the ion in its secular trap. This uncorrelated bosonic state does not prevent the occurrence of correlations between the ion and the bosonic ensemble, which are the key ingredients for the observation of entanglement between the two subsystems. In this way, as we shall see later, we can include thermal fluctuations in the degenerate Bose gas by employing the truncated Wigner (TW) method~\cite{SteelPRA98,SinatraPRL01,CockburnPRA11}. 

This paper is organized as follows. In Sec.~\ref{sec:frame} we outline the general theoretical model, while in Sec.~\ref{sec:spinEnta} we investigate the problem of many weakly interacting bosons for cases interacting with a static and a moving ion. Here the goal is to investigate the impact of ion motion and Bose gas finite temperature on the ion-controlled bosonic Josephson-junction dynamics. In Sec.~\ref{sec:motionEnta} we investigate entanglement generation between the ion and BJJ motional states, i.e., without an ion internal state. Finally, in Sec.~\ref{sec:outlook} we draw our conclusions and briefly discuss future research directions. 


\section{Theoretical framework}
\label{sec:frame}

In this section we introduce the microscopic many-particle Hamiltonian, together with the interactions involved in the problem, as well as the corresponding energy scales. Then, we outline the entanglement protocol we would like to implement that we shall investigate against ionic motion and finite temperature of the Bose gas. All this will form the basis for the subsequent analyses. 


\subsection{Hamiltonian}

\begin{figure}
\includegraphics[scale=0.6]{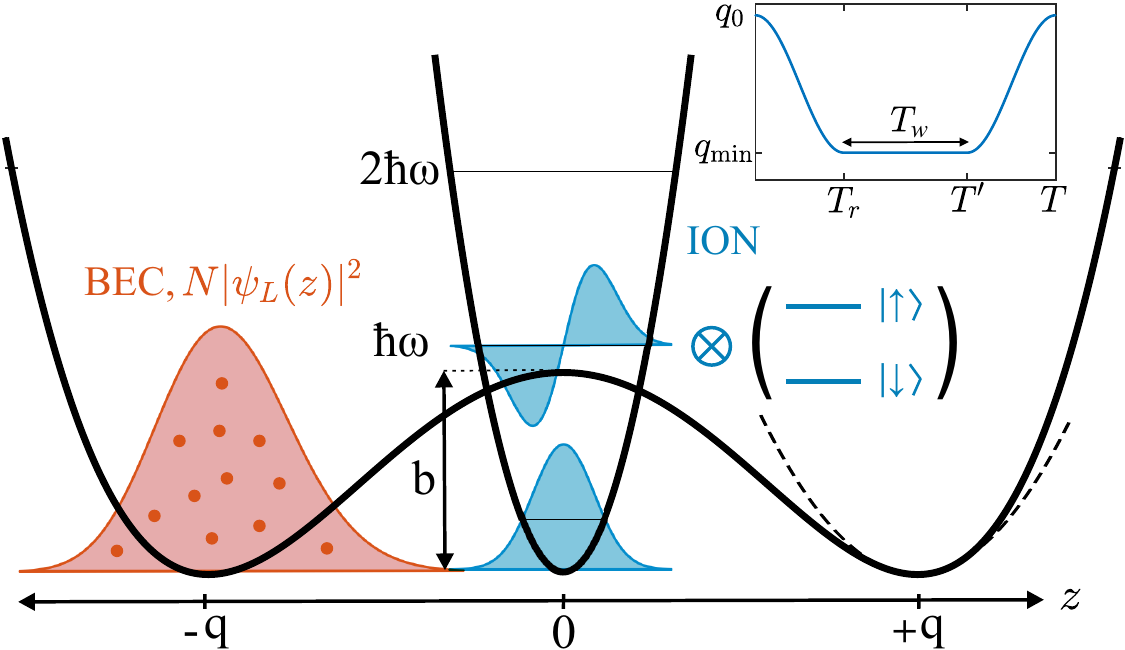}
\caption{\label{fig:f1} An ultracold quantum Bose gas, the atomic density $N\vert\psi_L(z)\vert^2$ of which is depicted by the shadowed red area on the left well, is confined in a double-well potential in interaction with a single ion confined in a harmonic trap of frequency $\omega$ located at the position of the barrier, i.e., $z=0$ (the ion energy levels are shifted by $\hbar\omega/2$). The dashed black line in the right well represents the harmonic approximation to the double-well potential, the frequency of which is denoted by $\omega_0$ (see main text). The ion internal spin (two-level system in parentheses) or vibrational states (shadowed blue functions) can be exploited to control the density current through the barrier. The symbol $\otimes$ indicates the ion Hilbert-space structure consisting of the motional and spin degrees of freedom. The entire setup constitutes an ion-controlled BJJ. The inset in the top right corner displays the transport function $q(t)$ given in Eq.~(\ref{eq:qt}).}
\end{figure}

We consider a trapped ion and an ensemble of bosonic atoms confined in a double well, such that both subsystems can be treated in one dimension (1D), i.e., their transverse motions are frozen to the corresponding ground state of their traps, as shown pictorially in Fig.~\ref{fig:f1}. Hence, the full Hamiltonian of the compound quantum system at ultralow temperatures is given by 
\begin{align}
\label{eq:Hfull}
\hat H & =  \sum_{j=1}^N \left[
\frac{\hat p^2_j}{2 m} + V_{\mathrm{dw}}(z_j) + \sum_{\xi=\downarrow,\uparrow}V_{\mathrm{ai}}^{\xi}(z_j-Z)\ket{\xi}\bra{\xi}
\right] \nonumber\\
\phantom{=} & + g\sum_{j < i}\delta(z_j - z_i) + \frac{\hat P^2}{2 M} + \frac{M\omega^2}{2}  Z^2.
\end{align}
Here $\hat P$ ($\hat p_j$) and $Z$ ($z_j$) denote the momentum operator and the spatial coordinate of the ion ($j$-th atom), and 
$\omega$ is the ion secular trap frequency, if the ion is confined in a Paul trap~\cite{LeibfriedRMP03} or the frequency of an optical trap~\cite{Lambrecht2016,SchaetzJPB17,SchmidtPRX18}. $M$ ($m$) denotes the mass of the ion (atomic boson) and $g = 2\hbar\omega_\perp a_s/(1 - 1.4603\, a_s/a_\perp)$ is the effective atom-atom interaction in a waveguide~\cite{OlshaniiPRL98}. Here $\omega_\perp$ is the frequency of the atom transverse trap [typically a few tens of $\omega_0$ -- 
see also below Eq.~(\ref{eq:dw})], $a_\perp = \sqrt{2\hbar/(m\omega_\perp)}$ is the width of the atomic waveguide, and $a_s$ is the three-dimensional (3D) $s$-wave atom-atom scattering length. 

For the double-well potential we consider the following simple analytical expression~\cite{MilburnPRA97} 
\begin{equation}
V_{\mathrm{dw}}(z)=\dfrac{b}{q^{4}}(z^{2}-q^{2})^{2},
\label{eq:dw}
\end{equation}
where $b$ denotes the inter-well barrier height and $2q$ is the distance between the two minima of the double-well potential (see also Fig.~\ref{fig:f1}).  At the minima, i.e., at $z=\pm q$, by Taylor expanding to second order the above outlined expression for the double-well potential, the potential can be approximated by a harmonic potential with frequency $\omega_{0}=\sqrt{8b/mq^{2}}$ (see dashed line in Fig.~\ref{fig:f1}). 

Finally, the atom-ion polarization potential is given by 
\begin{align}
\label{eq:Vai}
V_{\mathrm{ai}}(z - Z)=-C_{4}/(z - Z)^{4} 
\end{align}
with $C_{4}=\alpha e^{2}/{2}$~\footnote{We note that such form of the interaction is valid for atom-ion separations larger than $R_{\mathrm{1D}} = \max(\ell_\perp,R_\perp)$ with $\ell_\perp$ being the width of the atomic waveguide and $R_\perp$ the distance at which the 3D polarization potential becomes equal to the trapping potential in the transverse direction. For more details, we refer to~\cite{IdziaszekPRA07}}. Here $\alpha$ is the static polarizability of the atom and $e$ is the ion electric charge. Equation~(\ref{eq:Vai}) represents the atom-ion interaction at long-range distances, thus not depending on the internal state of the ion. However, at short-range, typically below a few nanometers, the atom and ion electronic configurations matter. Since such a reliance on the internal state is generally not known, as accurate energy potentials are extremely difficult to obtain, its impact on the atom-ion wave function is accounted for by so-called scattering short-range phase shifts, which are used as boundary conditions for solving the stationary Schr\"odinger equation~\cite{IdziaszekPRA07}. In this sense, the state-dependence of the atom-ion potential~(\ref{eq:Vai}) on the electronic configurations has to be understood. Furthermore, the interaction~(\ref{eq:Vai}) is characterized by typical energy and length scales that are denoted as: $R^{*}=\sqrt{\alpha e^{2}\mu/\hbar^{2}} $ and $E^{*}=[\hbar^{2}/2\mu (R^*)^2]$ with $\mu$ being the reduced mass. While one could employ quantum defect theory~\cite{IdziaszekPRA07} for solving both the time-independent and time-dependent Schr\"odinger equation, for the sake of numerical convenience we shall use the following model potential replacing~(\ref{eq:Vai}): 
\begin{equation}
V_{\mathrm{ai}}^{\xi}(z - Z)=v_{0}^\xi e^{-\gamma^\xi (z - Z)^{2}} - \frac{1}{(z - Z)^{4}+1/\varpi^\xi},
\label{eq:eq2}
\end{equation}
with $\xi = \downarrow,\,\uparrow$ denoting the internal spin state of the ion.
As discussed in detail in Ref.~\cite{SchurerPRA14}, the parameter $v_0^\xi$ is fixed to a relatively large number (in units of $E^*$) in order to ensure that the atom wave function at the ion location is close to zero. On the other hand, the parameters $\gamma^\xi$ and $\varpi^\xi$ are chosen in such a way that the scattering phase shifts at large distances of the one-dimensional (1D) atom-ion scattering problem at zero energy map to the corresponding ones obtained with quantum defect theory, from which the reliance of the short-range phases, $\varphi_{e}$ and $ \varphi_{o}$, on the model potential parameters is determined. Hence, a certain pair of model parameters $(\gamma^\xi,\varpi^\xi)$ does correspond to a specific pair of short-range phases $(\varphi_{e},\varphi_{o})$, thus to a particular ion spin configuration $\xi=\uparrow,\downarrow$. Specifically, as model parameters we have chosen~\cite{SchurerPRA16} $\varpi^{\downarrow}=29(R^{*})^{-4}$ and $\gamma^{\downarrow}=10\gamma_{\min}$ [$\varpi^{\uparrow}=80(R^{*})^{-4}$ and $\gamma^{\uparrow}=\gamma_{\min}$] with $\gamma_{\min}=4\sqrt{10\varpi}$, which correspond to the quantum defect parameters $\varphi_{e}^{\downarrow}=0.23\pi, \varphi_{o}^{\downarrow}=-0.45\pi$ ($\varphi_{e}^{\uparrow}=0.23\pi, \varphi_{o}^{\uparrow}=0.3\pi$). In both cases, we have set $v_{0}^\xi = 3\varpi^\xi$. 
We note that the short-range phases are related to the corresponding even and odd 1D atom-ion scattering lengths $a_{e,o}^{\uparrow,\downarrow} = - R^* \cot(\varphi_{e,o}^{\uparrow,\downarrow})$. 
These can be tuned either by means of magnetic Feshbach resonances or by the transverse external potential of the atom and the ion~\cite{IdziaszekPRA07,IdziaszekPRA09,MelezhikPRA16}. In the case where the atom-ion entanglement is controlled by the ion motion (see Sec.~\ref{sec:motionEnta}), we have used only the pair corresponding to the spin state $\ket{\downarrow}$. 

To conclude, we rescale the Hamiltonian in the following units: $\bar{R}^{*}=\sqrt{\alpha e^{2}m/\hbar^{2}}$ and $\bar{E}^{*}=[\hbar^{2}/2m (\bar{R}^*)^2]$, where we have replaced the reduced mass with the mass of the boson. For instance, for $^{87}$Rb we have $\bar{R}^{*}\simeq 375.31$ nm and $\bar{E}^{*}/h \simeq 0.41$ kHz. We prefer to make this choice, especially for the numerical simulations of the static ion case, since in this way the factor $\mu/m$ that otherwise would appear in the kinetic and trapping energies is removed. 


\subsection{Entanglement protocol}

Here we explain first which kind of entangled states we are aiming at and secondly how we can generate them.

\subsubsection{Target entangled states}

Our objective is to generate entanglement between either the internal or the motional state of the ion and the motional state of the atomic ensemble. In particular, we are interested in quantum states of the kind
\begin{align}
\label{eq:goalstate}
\ket{\Xi_{\mathrm{target}}} = c_0 \ket{\psi_L} \ket{\eta_0} + c_1 \ket{\psi_R} \ket{\eta_1},
\end{align}
where $c_{0,1}\in\mathbb{C}$ with $\vert c_{0}\vert^2+\vert c_1\vert^2 = 1$, and $\ket{\eta_n}$ with $n\in\mathbb{N}$ denotes either the eigenstate of the ion harmonic trap, i.e., the usual harmonic oscillator state ($n=0$ the ground state, $n=1$ the first excited state, etc.), or the ion internal state $\eta_1 \equiv \uparrow,\,\eta_0 \equiv \downarrow$. Also, the atomic states $\ket{\psi_{L,R}}$ with $\langle\psi_{L,R}\ket{\psi_{L,R}} = 1$ denote the states of the left and right well of the interacting bosonic ensemble, respectively. These states are linear combinations of the lowest energy symmetric and antisymmetric states of the double-well potential. In the many bosons case, such states are obtained by imaginary time propagation of the Gross-Pitaevskii equation (GPE)~\cite{EdwardsPRA95,DalfovoPRA96,ChiofaloPRE00}. In trapped ion experiments, superposition states $c_0 \ket{\eta_0} + c_1 \ket{\eta_1}$ of internal states can be generated by controlling their detuning with respect to the atomic transition and the light-ion interaction strength, i.e., Rabi frequency. On the other hand, ionic motional state superpositions can be attained by laser pulses in Raman configuration~\cite{LeibfriedRMP03}. To this end, at the initial time, the ion and the atomic ensemble are assumed to be noninteracting, which is attained by choosing a sufficiently large inter-well separation $q$ or, alternatively, a large barrier height $b$. Here, however, we choose to control dynamically $q$. In this situation, assuming that the atomic bosons are initially prepared in the left well (see also Fig.~\ref{fig:f1}), the initial quantum state of the compound system is simply the tensor product 
\begin{align}
\label{eq:initial}
\ket{\Xi_{\mathrm{initial}}}  = \ket{\psi_L} \otimes (c_0 \ket{\eta_0} + c_1 \ket{\eta_1}).
\end{align}
By properly choosing the model potential parameters as well as a tailored time dependence of the inter-well separation $q(t)$, it is possible, as we shall see later, to produce entangled states of the form~(\ref{eq:goalstate}). Because of the linearity of the Schr\"odinger equation (no spin-nonconserving interactions are assumed), however, we can focus simply on two separated processes, namely either $c_0= 0$ or $c_1=0$. 
Indeed, starting from the Hamiltonian~(\ref{eq:Hfull}) and from the most general representation of the many-boson--ion wave function
\begin{align}
\label{eq:Xit}
\vert\Xi(t)\rangle=c_{\uparrow}\vert\Psi_{\uparrow}(t)\rangle\vert\uparrow\rangle 
+ c_{\downarrow}\vert\Psi_{\downarrow}(t)\rangle\vert\downarrow\rangle,
\end{align}
with $\eta=\uparrow,\,\downarrow$ and $\vert\Psi_{\eta}(t)\rangle$ describing the motional state of both the ion and the bosonic ensemble when the ionic internal state is $\vert\eta\rangle$, one can easily verify that the full many-boson--ion Schr\"odinger equation reduces to two independent Schr\"odinger equations for each of the ion internal states $\vert\eta\rangle$. This is a consequence of the fact that we assume no dynamics for the ion internal state, e.g., due to spin-exchange collisions, as indeed the Hamiltonian~(\ref{eq:Hfull}) ensures. We then use a different ansatz for the wave function $\vert\Psi_{\eta}(t)\rangle$ depending on whether we treat the ion statically or not, as we shall discuss later in the paper. Hence, if we can identify suitable short-range atom-ion scattering parameters together with a proper choice of the function $q(t)$, such that the two processes
\begin{align}
\label{eq:gate}
\ket{\psi_L} \ket{\eta_0} \mapsto \ket{\psi_R} \ket{\eta_0} \qquad 
\ket{\psi_L} \ket{\eta_1} \mapsto \ket{\psi_L} \ket{\eta_1}
\end{align}
can be realized efficiently, we can indeed generate any superposition state of the kind~(\ref{eq:goalstate}). This approach is analogous to the typical strategy employed for verifying the validity of an implementation of an entangled two-qubit gate with two atoms~\cite{CironeEJPD05,TreutleinPRA06}; rather than showing what happens to a particular superposition, one simply demonstrates the feasibility of the entanglement scheme on the computational basis and then for linearity it applies to any superposition. We note that the transformation~(\ref{eq:gate}) is indeed equivalent to a controlled-NOT two-qubit gate, as we pointed out in Ref.~\cite{SchurerPRA16}, where the internal or motional state of the ion plays the role of the control qubit. We do not aim at investigating the performance of a new scheme for a two-qubit quantum gate, however, but rather at demonstrating the feasibility to generate with such a compound quantum system mesoscopic cat-like states, as the ones between a light field and a single Rydberg-atom~\cite{HarocheRMP13}, against relevant experimental imperfections. In conclusion, hereafter, when we refer to atom-ion entanglement generation, we essentially refer to the attainment of the transformations~(\ref{eq:gate}).

\subsubsection{Protocol}

As already pointed out, the two aforementioned dynamical processes are controlled by the inter-well separation, for which we have chosen the particular form: 
\begin{equation}
\label{eq:qt}
q(t)=\left\{
                \begin{array}{ll}
                  q_{0} \hspace{5.8cm} t<0 \\
                  \\
                  \dfrac{q_{0}-q_{\min}}{2}f(t)+\dfrac{q_{0}+q_{\min}}{2} \hspace{1.4cm} 0 \le t < T_{\mathrm{r}} \\
                  \\
                  q_{\min} \hspace{4.6cm} T_{\mathrm{r}} \le t < T^{\prime}, \\
                  \\
                  \dfrac{q_{0}-q_{\min}}{2}f(t-T)+\dfrac{q_{0}+q_{\min}}{2} \hspace{0.5cm} T^{\prime} \le t \le T \\
                  \\
                  q_{0} \hspace{5.8cm}  t>T \\
                \end{array}
              \right.
\end{equation} 
where $f(t)=\cos(\pi t/T_{\mathrm{r}})$, $T=2T_{\mathrm{r}}+T_{\mathrm{w}}$, and $T^{\prime}=T_{\mathrm{r}}+T_{\mathrm{w}}$ (see also inset in Fig.~\ref{fig:f1}). Here $T_{\mathrm{r}}$ is the time for decreasing and increasing the inter-well separation, whereas  $T_{\mathrm{w}}$ denotes the ``waiting time'', i.e., the time needed to transport the bosons from one well to the other one. Furthermore, the maximum inter-well separation is denoted by $q_{0}$, whereas $q_{\min}$ is the minimum inter-well separation reached in the dynamics. We note that hereafter we choose $q_{\min}$ in 
such a way that it is larger than the critical atom-ion separation $q_c=2 [R^*\hbar / (M\omega)]^{1/3}$ below which the effects of micromotion become significantly detrimental \cite{NguyenPRA12,JogerPRA14}. We note, however, that micromotion is not an issue for ion trapping techniques based on optical fields~\cite{Lambrecht2016,SchaetzJPB17,SchmidtPRX18} and that for linear Paul traps it is advisable to engineer the double-well potential along the longitudinal ion trap axis, where the radiofrequency fields are absent. As a rule of thumb, in order to identify the most suitable parameters defining $q(t)$, we fix $T_{\mathrm{r}}=12 \hbar/\bar{E}^{*}$~\footnote{Hereafter $T_{\mathrm{r}}=12 \hbar/\bar{E}^{*}$, unless differently stated.}, $q_{0}=5 \bar{R}^{*}$ and $b=5.5\bar{E}^{*}$, which ensure that initially the Bose gas and the ion do not interact, while $T_{\mathrm{w}}$ varies within the range $10 - 90\hbar/\bar{E}^{*}$ and $q_{\min}$ varies within the range $2 - 3\bar{R}^{*}$ (experimentally realistic parameters can be found in Ref~\footnote{
The parameters quoted above would correspond to $(q_0 \simeq 383.25, q_{\min} \simeq 153.30 - 229.95)$ nm, $(T_{\mathrm{r}} \simeq 0.02, T_{\mathrm{w}} \simeq 0.01 - 0.12)$ ms for the atom-ion pair $^7$Li / $^{174}$Yb$^+$, and $(q_0 \simeq 1.88, q_{\min} = 0.75 - 1.13)$ $\mu$m $(T_{\mathrm{r}} \simeq 4.65, T_{\mathrm{w}} \simeq 3.87 - 34.86)$ ms for $^{87}$Rb / $^{171}$Yb$^+$, which are experimentally realistic~\cite{TreutleinPRA06,LesanovskyPRA06,HofferberthNP06,MaussangPRL10,BetzPRL11,ValtolinaScience15,PigneurPRL18,BurchiantiPRL18}. On the other hand, for the critical atom-ion separation, assuming $\omega=\omega_0$, we have for example: $q_c = 169.67$ nm for $\omega= 2\pi\times 7.27$ kHz and $^7$Li / $^{174}$Yb$^+$, whereas $q_c = 572.83$ nm for $\omega= 2\pi\times 0.77$ kHz and $^{87}$Rb / $^{171}$Yb$^+$.}). In this way, the fidelity, defined in Eq.~(\ref{eq:fidelity}), of the transformations~(\ref{eq:gate}) at final time $T$ is larger than 95\%. 

Finally, the performance of the processes~(\ref{eq:gate}) is assessed by computing the Uhlmann fidelity defined as~\cite{UhlmannRMP76,Nielsen00} 
\begin{equation}
\label{eq:fidelity}
F(t)=\mathrm{Tr}[\sqrt{\sqrt{\hat{\rho}_{G}}\hat{\sigma}(t) \sqrt{\hat{\rho}_{G}}}].
\end{equation}
Here $\hat{\rho}_{G}$ is the density matrix of the goal state we aim at, i.e., $\hat{\rho}_{G}=\ket{\psi_{L,R}}\bra{\psi_{L,R}}$ for the bosonic system or $\hat{\rho}_{G}=\ket{n=0,1}\bra{n=0,1}$ for the ion motion, whereas $\hat{\sigma}(t)$ is the reduced density matrix of either the bosonic or the ionic system at time $t$ obtained by simulating the many-body quantum dynamics with the protocol outlined above. Note that the above goal density matrices $\hat{\rho}_{G}$ do correspond to pure states. Thus, attaining unity fidelity for these states implies that the bosonic state is one of the two pure states $\ket{\psi_{L,R}}$. In particular, we use the one-body density matrix of the quantum state of the Bose gas. We also note that for pure states the Uhlmann fidelity reduces to the absolute value of the overlap integral between the evolved state $\ket{\psi(t)}$ and the goal state, that is
\begin{equation}
\label{eq:overlaps}
F_{L,R}(t)=\vert\langle\psi_{L,R}\vert\psi(t)\rangle\vert.
\end{equation}
We underscore that the Uhlmann fidelity, which yields a real number between zero and unity, is not directly a measure of the entanglement between the atomic ensemble and the ion, but rather a measure of the quality of the transformations~(\ref{eq:gate}). The higher the quality of such transformations is, i.e., the closer the Uhlmann fidelity to unity is, the higher the quality of the target state $\ket{\Xi_{\mathrm{target}}}$ is. For instance, assuming the ion to be static and the bosonic ensemble to be at zero temperature, a measure of the attainment of the desired target state is given by the overlap fidelity 
\begin{equation}
\label{eq:Ftarget}
\vert \langle\Xi_{\mathrm{target}}\vert\Xi(T)\rangle\vert \le \vert c_\uparrow\vert^2 F_\uparrow(T) + \vert c_\downarrow\vert^2 F_\downarrow (T) \le 1,
\end{equation}
where $F_\uparrow(T)\equiv F_L(T)$, $F_\downarrow(T)\equiv F_R(T)$, and $\vert\Xi(T)\rangle$ given by Eq.~(\ref{eq:Xit}), but in mean field. Hence, in our subsequent analyses we focus on attaining values of the Uhlmann fidelities for the two mentioned processes as close as possible to unity, rather than concentrating our attention on a specific choice of the coefficients $c_\uparrow$ and $c_\downarrow$, i.e., on a particular entangled state. Indeed, in order to accurately assess the overlap $\vert \langle\Xi_{\mathrm{target}}\vert\Xi(T)\rangle\vert$ one would also need to investigate the performance of the protocol in generating the initial superposition~(\ref{eq:initial}) of internal or motional states of the ion, which is beyond the scope of the present paper.


\section{Entanglement generation with the ion internal state}
\label{sec:spinEnta}

In this section we investigate the impact of finite temperature of the degenerate Bose gas and the ion motion on the entanglement protocol when 
the ion internal degree of freedom controls the tunneling of the bosons through the barrier, as suggested in Refs.~\cite{GerritsmaPRL12,SchurerPRA16}. 
Hence, here we assume that the ion is prepared, and possibly kept, in the ground state of its (secular) trap. 


\subsection{Zero temperature}
\label{sec:Tzero}

To begin with, we consider the case of a static ion. In this scenario, the dynamics is well described by the GPE
\begin{equation}
\label{eq:GPE}
\begin{aligned}
i \hbar \dfrac{\partial \psi}{\partial t} = \Bigg[\dfrac{\hat p^{2}}{2m}+V_{\mathrm{dw}}(z)+V_{\mathrm{ai}}^\xi(z)+g(N-1)\vert\psi\vert^{2}\Bigg]\psi,
\end{aligned}
\end{equation} 
where $\psi=\psi(z,t)$ is the condensate wave function normalized to unity, $N=50$, and $g=0.004\, \bar{E}^* \bar{R}^*$. 
At the mean-field level, the tunneling and the self-trapping processes can be characterized by the single parameter~\cite{SmerziPRL97}
\begin{equation}
\label{eq:Lambda}
 \Lambda=\dfrac{UN}{2J},
\end{equation}  
where $U=g \int \vert \psi_{L} (z) \vert ^{4} dz$ is the onsite energy and $J/\hbar = \frac{1}{\hbar} \int \psi_{L}^* (z) \left(\dfrac{\hat p^{2}}{2m}+V_{\mathrm{dw}}(z)+V_{\mathrm{ai}}^\xi(z)\right) \psi_{R}(z)  dz$ is the tunneling rate. Upon the initial conditions, one can identify a critical value of $\Lambda$ above which self-trapping occurs. For instance, for the trap parameters $b=5.5 \bar{E}^{*}$ and $q_{\min}=2.55 \bar{R}^{*}$ we have $\Lambda_{c}=2$. Particularly with the ion, we have $\Lambda=0.54$ for $(\varpi^{\downarrow},\gamma^{\downarrow})$, i.e., tunneling, and $\Lambda=5.97$ for $(\varpi^{\uparrow},\gamma^{\uparrow})$, i.e., the BEC is self-trapped. This confirms that the BJJ can be indeed controlled by a single trapped ion. 

\begin{figure}
\includegraphics[scale=0.44]{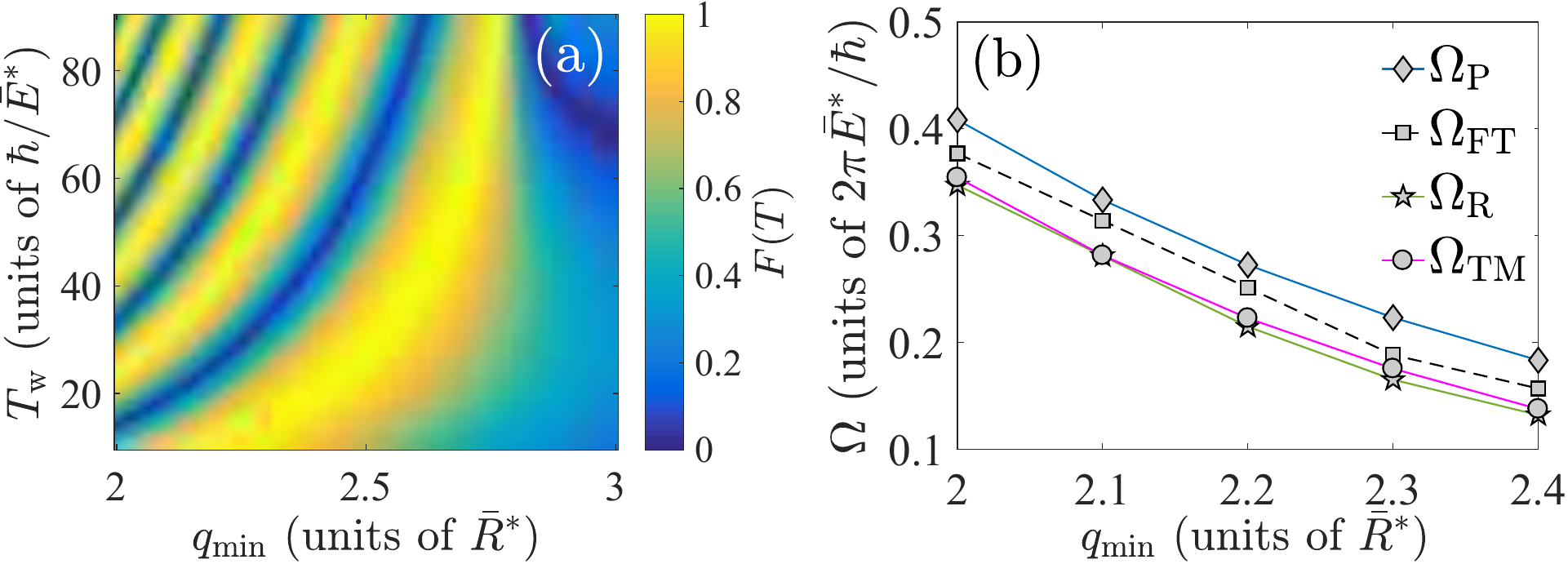}
\caption{\label{fig:f7} 
(a) Uhlmann fidelity at the final time $T$ for a BEC interacting with a static ion as a function of the waiting time $T_{\mathrm{w}}$ and minimal well separation from the barrier $q_{\min}$ for the model parameter pair $(\varpi^{\downarrow},\gamma^{\downarrow})$, which corresponds to the tunneling regime. The barrier height is set to $b= 5.5 \bar{E}^*$.
(b) Oscillation frequency of the Uhlmann fidelity shown in panel (a), i.e., along its vertical axis, for various minimal separations $q_{\min}$: $\Omega_{\mathrm{P}}$ is the plasma frequency, $\Omega_{\mathrm{FT}}$ is the frequency of the oscillations of panel (a) obtained via Fourier transform, $\Omega_{\mathrm{R}}$ is the frequency of the Rabi weakly interacting regime, and $\Omega_{\mathrm{TM}}$ is the frequency obtained with the two-mode mean-field model. The continuous lines are merely a guide to the eye.
}
\end{figure}

\begin{figure}
\includegraphics[scale=0.48]{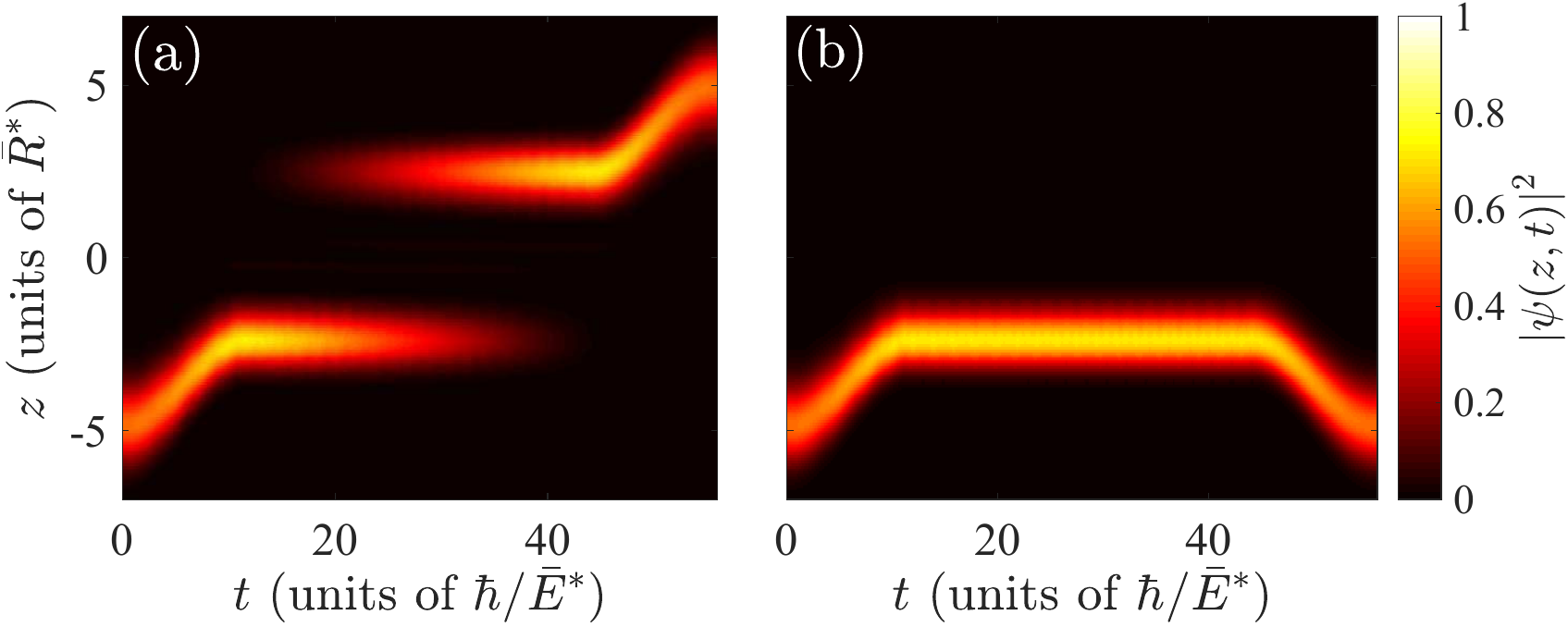}
\caption{\label{fig:f9} Temporal evolution of the atomic density (normalized to unity) for the pair $(\varpi^{\downarrow},\gamma^{\downarrow})$ (a) and for the pair $(\varpi^{\uparrow},\gamma^{\uparrow})$ (b) during the entanglement process for a BEC and a static ion. Here, $q_{\min}=2.55\bar{R}^{*}$, $b=5.5\bar{E}^*$, and $T_{\mathrm{w}}=32\hbar/\bar{E}^{*}$, which correspond to $\omega_0 \simeq 2\pi \times7.27$ kHz for $^7$Li atoms and to $\omega_0 \simeq 2\pi \times0.77$ kHz for $^{87}$Rb atoms. The frequency $\omega_0$ is computed at $q=q_0$, that is, when initially atom and ion do not interact.}
\end{figure}

We have analyzed the performance of the protocol by computing the Uhlmann fidelity by varying $T_{\mathrm{w}}$ and $q_{\min}$, but for fixed $T_{\mathrm{r}}=12 \hbar/\bar{E}^{*}$ and $q_{0}=5 \bar{R}^{*}$. The result of this investigation is summarized in Fig.~\ref{fig:f7} (a) for the tunneling process only, since for the other pair of model potential parameters, regardless of $T_{\mathrm{w}}$ and $q_{\min}$, the bosonic ensemble remains essentially in its original starting well, namely, we have always macroscopic self-trapping. As it can be seen, there are revivals dependent on the waiting time and the minimal separation between the wells. This indicates that there is a certain flexibility in the choice of the trapping and timing parameters. In order to gain insight about the origin of the revivals along the waiting time $T_{\mathrm{w}}$ at fixed $q_{\min}$, however, we have performed a Fourier analysis which is summarized in Fig.~\ref{fig:f7} (b). We compared the numerically extracted oscillation frequency via Fourier transform, $\Omega_{\mathrm{FT}}$, with the one obtained within the mean-field two-mode approximation~\cite{SmerziPRL97,RaghavanPRA99}, $\Omega_{\mathrm{TM}}$, the plasma frequency $\Omega_{\mathrm{P}} = \sqrt{2U N J + 4 J^2}$~\cite{SmerziPRL97}, where both $U$ and $J$ rely on $q_{\min}$, and with the Rabi frequency $\Omega_{\mathrm{R}} = 2 J/\hbar$. As it can be seen, since the processes simulated in Fig.~\ref{fig:f7} (a) are not precisely adiabatic, we cannot attribute the oscillations of $F$ to a single frequency mode. Nonetheless, we note that for large values of $q_{\min}$, $\Omega_{\mathrm{TM}}$ -- or $\Omega_{\mathrm{R}}$ as they are very similar -- is the dominating one, whereas for small values of $q_{\min}$ the plasma frequency $\Omega_{\mathrm{P}}$ seems to prevail over the other two modes. To conclude this analysis, in Fig.~\ref{fig:f9} we show, as an example, the (normalized) density evolution of the two processes, i.e., tunneling and self-trapping, the fidelity of which is in both cases larger than 99\%, thus proving the feasibility of the protocol. 

Now, the natural question is the following: How does the ion motion influence the BJJ dynamics and specifically the transformations~(\ref{eq:gate}) of the ion internal degree of freedom and the motional state of the BEC? In order to answer this question, we first have to define a proper ansatz for the correlated BEC-ion wave function and derive equations of motion that take into account the ion motion coupled to the quantum Bose gas. Towards that end, we introduce the ansatz
\begin{equation}
\label{eq:Psit}
\vert\Psi(t)\rangle=\sum_{n=0}^{n_{i}} c_{n}(t)\vert\Phi_{n}(t)\rangle\vert\phi_{n}\rangle,
\end{equation}
where $\vert\Phi_{n}(t)\rangle$ denotes the wave function of $N$ bosons when the ion state is in the $n$-th harmonic oscillator state $\vert\phi_{n}\rangle$, whereas $\sum_{n=0}^{n_{i}}\vert c_{n}(t)\vert^{2}=1$ with $n_i$ being a cutoff, i.e., we truncate the ion Hilbert space by considering its most relevant portion (typically $n_i=3$ is sufficient for the dynamics treated in this paper, as also discussed later). The complex numbers $c_{n}$ reveal the degree of  entanglement between the degenerate Bose gas and the ion (i.e., the Schmidt number~\cite{Nielsen00}), namely if $c_{0}=1$ and $c_{n\neq 0}=0$, then the compound quantum system is not entangled. In general, solving the dynamics of the compound quantum system for time-dependent $c_n$ and $\vert\Phi_{n}\rangle$ can be very involved. The previous study~\cite{SchurerPRA16} on the static ion approximation shows that it is very reasonable to make the ansatz $\lvert \Phi_{n}(t)\rangle=\prod_{j=1}^{N}\lvert \varphi_{n}^{(j)}(t)\rangle$, namely assume no correlations between the bosons, at least for times shorter than a few tens of $\hbar/\bar{E}^*$ like in Fig.~\ref{fig:f9}. Thus, by employing the so-called Dirac-Frenkel variational principle~\cite{DiracMPCPS30,Frenkel34}, we can derive the corresponding equations of motion for $c_n(t)$ and $\lvert \varphi_{n}^{(j)}(t)\rangle$, which are outlined in Eqs.~(\ref{eq:eq19}-\ref{eq:eq20}) in the appendix. By solving these equations, we can then obtain the corresponding one-body density matrix for the bosons as well as the density matrix of the ion from the full density matrix $\hat \rho = \vert\Psi\rangle\langle \Psi\vert$. The $N$-body density matrix is given by 
\begin{align}
\label{eq:rBG}
\hat \rho_{\mathrm{BG}} & = \mathrm{Tr}_{\mathrm I}(\hat \rho) = \sum_{n} \vert c_n \vert^2 \hat \rho_{\mathrm{BG}}^{(n)},
\end{align}
where $\rho_{\mathrm{BG}}^{(n)} = \vert \Phi_{n}\rangle\langle\Phi_n\vert$ is the density matrix of the bosonic ensemble when the ion is in the $n$-th motional state of the trap. Thus, $\hat \rho_{\mathrm{BG}}$ is the weighted average over the motional states of the ion. The one-body density matrix of the degenerate Bose gas is then obtained by tracing out $N-1$ bosons, which yields
\begin{align}
\label{eq:rBG1body}
\hat \rho_{\mathrm{BG}}^{1} & = \mathrm{Tr}_{N-1}(\hat \rho_{\mathrm{BG}} ) = \sum_{n} \vert c_n \vert^2 \ket{\varphi_{n}}\bra{\varphi_{n}}.
\end{align}
On the other hand, the ion density matrix is given by 
\begin{align}
\label{eq:rI}
\hat\rho_{\mathrm{I}} = \mathrm{Tr}_{\mathrm{BG}}(\hat \rho) = \sum_{n,n^\prime=0}^{n_i} c_n c_{n^\prime}^* \langle\varphi_{n^\prime}\vert\varphi_{n}\rangle^N \ket{\phi_{n}}\bra{\phi_{n^\prime}}.
\end{align}
Hence, the fidelity of each of the two processes is determined by replacing $\hat\sigma$ in Eq.~(\ref{eq:fidelity}) with $\hat \rho_{\mathrm{BG}}^1$ of Eq.~(\ref{eq:rBG1body}) and by the ground state occupancy of the ion. In other words, the total fidelity is defined as 
\begin{align}
\label{eq:totalF}
\mathcal{F} = F \vert c_0\vert^2. 
\end{align}
We note that while the states $\ket{\varphi_{n}}$ are normalized to unity, they are in general not orthogonal to each other. 

\begin{figure}
\includegraphics[scale=0.45]{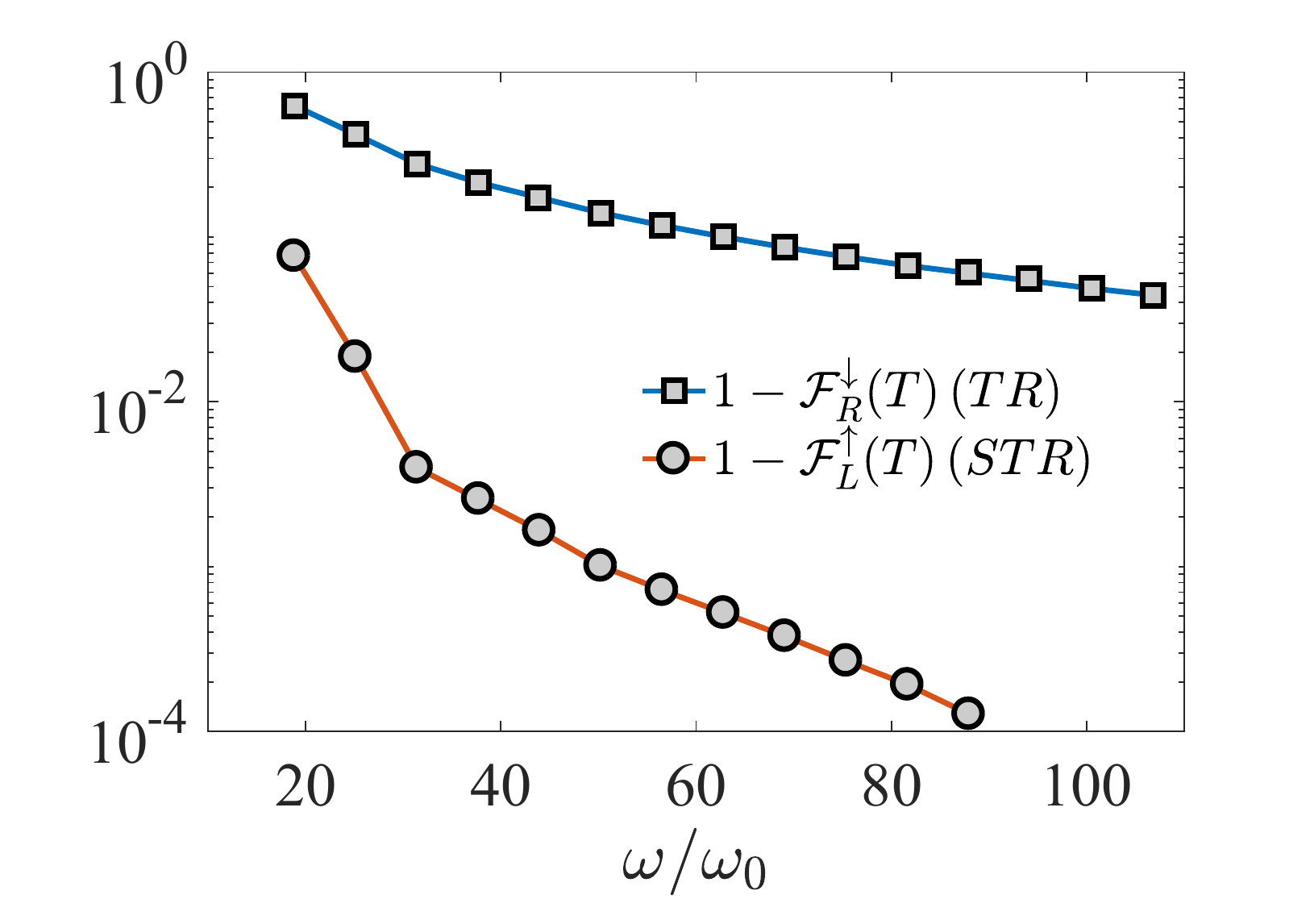}
\caption{\label{fig:f10} Total infidelity, i.e., 1 - $\mathcal{F}$, as a function of the ratio $\omega/\omega_0$ for the pair $(\varpi^{\downarrow},\gamma^{\downarrow})$ (squares) and for $(\varpi^{\uparrow},\gamma^{\uparrow})$ (circles). Here $q_{\min}=2.55\bar{R}^{*}$, $b=5.5\bar{E}^{*}$, and $T_{\mathrm{w}}=32\hbar/\bar{E}^{*}$. The continuous line is merely a guide to the eye.
} 
\end{figure}

We have analyzed the effect of the ion motion on the entanglement protocol for different ion trap frequencies. The findings are illustrated in Fig.~\ref{fig:f10}, where we have chosen to rescale the ion trap frequency in units of the double-well characteristic frequency $\omega_0$. We see that the larger the ion trap frequency is the better the static ion approximation is. In particular, for the self-trapping regime (STR) (circles) this holds already for moderated trap frequencies, i.e., larger than (approximately) 30$\,\omega_0$, whereas for the tunneling regime (TR) (squares) the ion trap has to be much tighter in order to attain the static ion limit. This can be explained by the fact that for shallower ion traps the ion wave function plays a major role in the tunneling dynamics of the bosons, because of the long-ranged atom-ion interaction. This effect is completely neglected in the static ion limit, where the reliance of the tunneling rate on the ion is only due to the atom-ion short-range phases, i.e., the short-range part of the atom-ion potential. On the other hand, for the STR it turns out that the $\Lambda$ parameter~(\ref{eq:Lambda}) is only marginally affected by the ion wave function, that is, it is still above $\Lambda_c$, thus ensuring the self-trapping condition, unless $\omega\le10\,\omega_0$ as we found from a numerical analysis. We note that in that scenario a Markovian description of the ion dynamics in the framework of quantum master equations~\cite{KrychPRA15} would not hold, as the correlation functions of the bosonic bath would decay on the time scale of the ion's dynamics.  

\begin{figure*}
\includegraphics[scale=0.95]{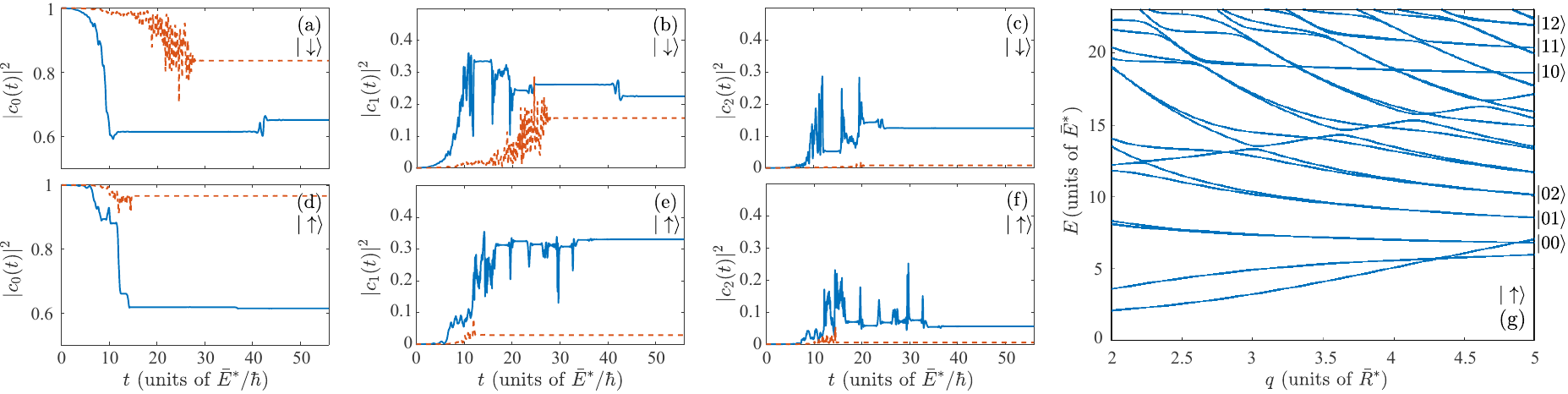}
\caption{\label{fig:f12} Time evolution of the occupancies $\vert c_n\vert^2$ for $n=0,1,2$ for the pair $(\varpi^{\downarrow},\gamma^{\downarrow})$, i.e., internal ion state $\ket{\downarrow}$ (a-c), and for $(\varpi^{\uparrow},\gamma^{\uparrow})$, i.e., internal ion state $\ket{\uparrow}$ (d-f). 
The trap parameters are chosen as in Fig. \ref{fig:f10}. The blue line corresponds to $\omega = 6.3 \,\omega_0$, whereas the red dashed line corresponds to $\omega = 12.6\, \omega_0$. In the panel (g) we show an example of the energy spectrum for an atom and an ion as a function of the separation $q$ for the model parameters corresponding to the ion internal state $\ket{\uparrow}$. The labels $\ket{n_i,n_a}$ on the right-hand side of the panel indicate the number of excited quanta in the ion ($n_i$) or in the atom ($n_a$) trap, respectively. In this case the ion trap frequency is $\omega = 6.3 \,\omega_0$. 
}
\end{figure*}
\begin{figure}
\includegraphics[scale=0.48]{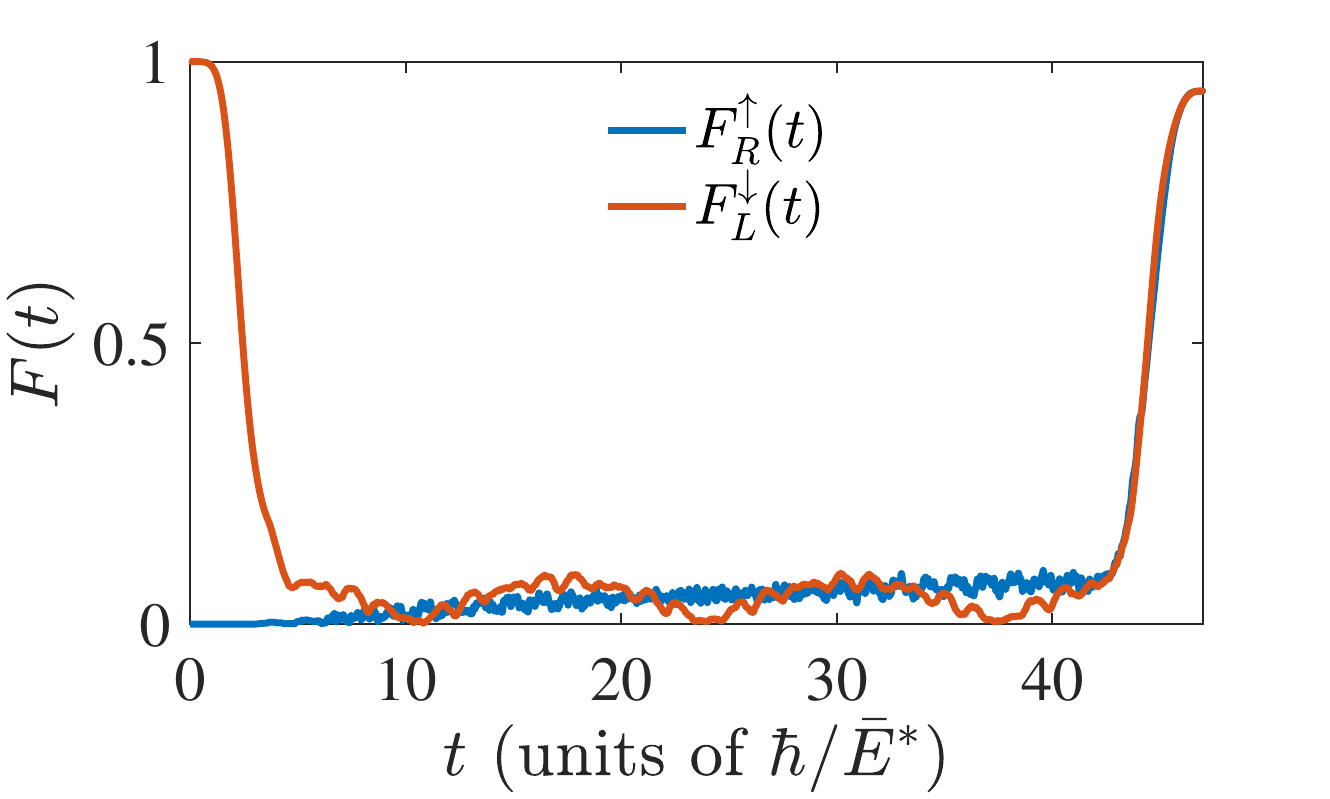}
\caption{\label{fig:f11} Uhlmann fidelity~(\ref{eq:fidelity}) with goal density matrices $\hat{\rho}_{G}=\ket{\psi_{L,R}}\bra{\psi_{L,R}}$ for a moving ion in a harmonic trap of frequency 
$\omega=18.8\omega_{0}$ and $q_{\min} = 2.05\,\bar{R}^*$, $q_{0} = 5\,\bar{R}^*$, $T_{\mathrm{w}} = 37\,\hbar/\bar{E}^*$, $T_{\mathrm{r}} = 5\,\hbar/\bar{E}^*$.}
\end{figure}

 
Finally, in Fig.~\ref{fig:f12} we display the time evolution of the occupancies $\vert c_n\vert^2$, that is, the population of the vibrational states of the ion trap [see also Eq.~(\ref{eq:Psit})], for two ion trap frequencies. 
First of all, we note the appearance of peaks for ion vibrational states higher than the ground state [see panel (b)-(c) and (e)-(f)]. We attribute these peaks to the occurrence of several avoided crossings in the single-atom and single-ion energy spectrum as a function of the well separation $q$, as shown in the panel (g) of Fig.~\ref{fig:f12} as well as in a detailed analysis in Ref.~\cite{JogerPRA14}. 
As it can be seen, the larger the ion trap frequency is, the smaller the contribution of energetically higher trap states is. As we already pointed out, for shallower ion traps, the ion wave function is broader and, because of the long-range atom-ion polarization potential, the interaction occurs over a longer time. Thus, not only does the ion modify the tunneling rate of the bosons, but also the atoms significantly affect the ion motion by causing population of additional vibrational states. Hence, the condition for which the ion is kept in the ground state is no longer satisfied, and, as a consequence, the total fidelity $\mathcal{F}$ decreases. Therefore, the static ion approximation is reasonably good for both processes if the ion trap frequency is roughly larger than 100$\,\omega_0$ (see also Fig.~\ref{fig:f10}), when we choose particular values of the trap and timing parameters. By properly choosing those parameters, however, as shown in Fig.~\ref{fig:f11} for a particular set of trap and timing parameters, we find that the processes~(\ref{eq:gate}) are still possible with quite good fidelities (above 90\%) even at shallow ion traps (cf. Fig.~\ref{fig:f10} for $\omega/\omega_0 = 18.8$). At the same time, the ion is almost in the ground state of its trap, i.e., $\min_t\{\vert c_0(t)\vert^2\} \ge 0.96$. We note, however, that in this case, since the ion trap frequency is smaller than $30\omega_0$ (see also the cusp in the red line of Fig.~\ref{fig:f10}), we have for both ion internal states the tunneling regime, but with commensurable rates. More specifically, for the state $\ket{\downarrow}$ (red line in Fig.~\ref{fig:f11}) the tunneling rate is twice that of the state $\ket{\uparrow}$ (blue line in Fig.~\ref{fig:f11}). Moreover, the goal density matrices are $\hat{\rho}_{G}^\downarrow=\ket{\psi_{L}}\bra{\psi_{L}}$ and $\hat{\rho}_{G}^\uparrow=\ket{\psi_{R}}\bra{\psi_{R}}$.

In summary, while the ion trap plays a non-negligible role in the Josephson dynamics, the entanglement protocol is attained efficiently by an accurate search of parameters. 

\subsection{Finite temperature}
\label{sec:temperature}

We proceed further with our analysis by investigating the impact of thermal fluctuations of the Bose gas on the correlated BEC-ion quantum state. To this end, we employ the truncated Wigner method~\cite{SteelPRA98}. This method, which belongs to the family of so-called classical fields methods~\cite{ProukakisJPB08,Proukakis2013}, consists of generating a stochastic ensemble of initial matter-wave fields that are then propagated by means of the time-dependent GPE. Contrary to the usual GPE, which describes the evolution of the condensate wave function only, in the TW method the wave function describes the entire matter-wave field, i.e., both the condensed and noncondensed part. By averaging over a sufficiently large sample of stochastic fields, one can compute quantities like one-body and two-body density matrices that can be then used to obtain observables of interest, e.g., correlation functions. Furthermore, in the TW method the temperature of the gas is set initially by choosing a thermal state at temperature $\mathcal{T}$. In the particle-number-conserving Bogoliubov approach~\cite{GardinerPRA97,CastinPRA98} that we adopt here, the thermal state is a canonical ensemble in which the many-body Hamiltonian is replaced by the quadratic Bogoliubov Hamiltonian~\cite{SinatraPRL01}. Within this approach the number of noncondensed atoms has to be small compared to $N$, thus implying that relatively low temperatures can be treated. At the same time $k_B\mathcal{T}/(\hbar\omega_0)> 1$ has to be fulfilled, that is, the temperature cannot be too low as well. For further details on the method, see Refs.~\cite{SinatraJPB02,CockburnPRA11}.

We begin our analysis by investigating the static ion limit. In this scenario, the standard TW method can be straightforwardly employed~\footnote{Here we use the number conserving Bogoliubov approach of Ref.~\cite{SinatraPRL01}.}, since the atom-ion polarization potential appears in the GPE~(\ref{eq:GPE}) as an additional external potential for the bosons. In this setting, as a goal state, i.e., the density matrix $\hat\rho_G$, we use the initial one-body density matrix of the Bose gas at temperature $\mathcal T$ when the two subsystems are well separated, i.e., the ion does not interact with the quantum gas, i.e.

\begin{align}
\label{eq:rhoG}
\rho_{G} (z,z^\prime) =  \langle z\vert \hat \rho_{G}\vert z^\prime \rangle= \langle \varphi_{0}^*(z) \varphi_{0}(z^\prime)\rangle_W - n_q \delta_{z,z^\prime}.
\end{align}
Here $n_q = 1/ (2 \Delta z)$ mimics the $\delta$-function on a spatial grid ($\Delta z$ is the grid spacing chosen in the numerics), $\delta_{z,z^\prime}$ is the Kronecker delta, and $\langle \varphi_{0}^*(z) \varphi_{0}(z^\prime)\rangle_W$ represents the statistical average over the $N_s$ stochastic fields distributed according to the Wigner representation of the $N$-body density matrix of the Bose gas~\footnote{Here $\varphi_{0}$ refers to the single-particle orbital when the ion is in the ground state of the trap. Of course, in the static ion limit this is the only state that can be populated.}. Note, however, that in this context the one-body density matrix is normalized to $N+\mathcal{M}/2$ with $\mathcal{M}$ being the number of Bogoliubov modes used in the expansion of the noncondensed matter field (see Refs.~\cite{SinatraPRL01,CockburnPRA11} for further technical details). Thus, in order to make a meaningful comparison with the zero-temperature case, we first normalize to unity the density matrices, e.g., $\rho_{G}(z,z^\prime)\mapsto \rho_{G} (z,z^\prime) / (N + \mathcal{M}/2)$. This state, however, differs from $\hat{\rho}_{G}=\ket{\psi_{L,R}}\bra{\psi_{L,R}}$. Nonetheless, we note that for the purpose of the entanglement generation we do not necessarily require being in the condensate mode, but in some motional state of the gas, e.g., a thermal state, localized in either the left or the right well. Finally, the time-evolved density matrix $\sigma(z,z^\prime,t=T)$ is determined similarly to $\rho_{G}(z,z^\prime)$. 

\begin{figure}
\includegraphics[scale=0.45]{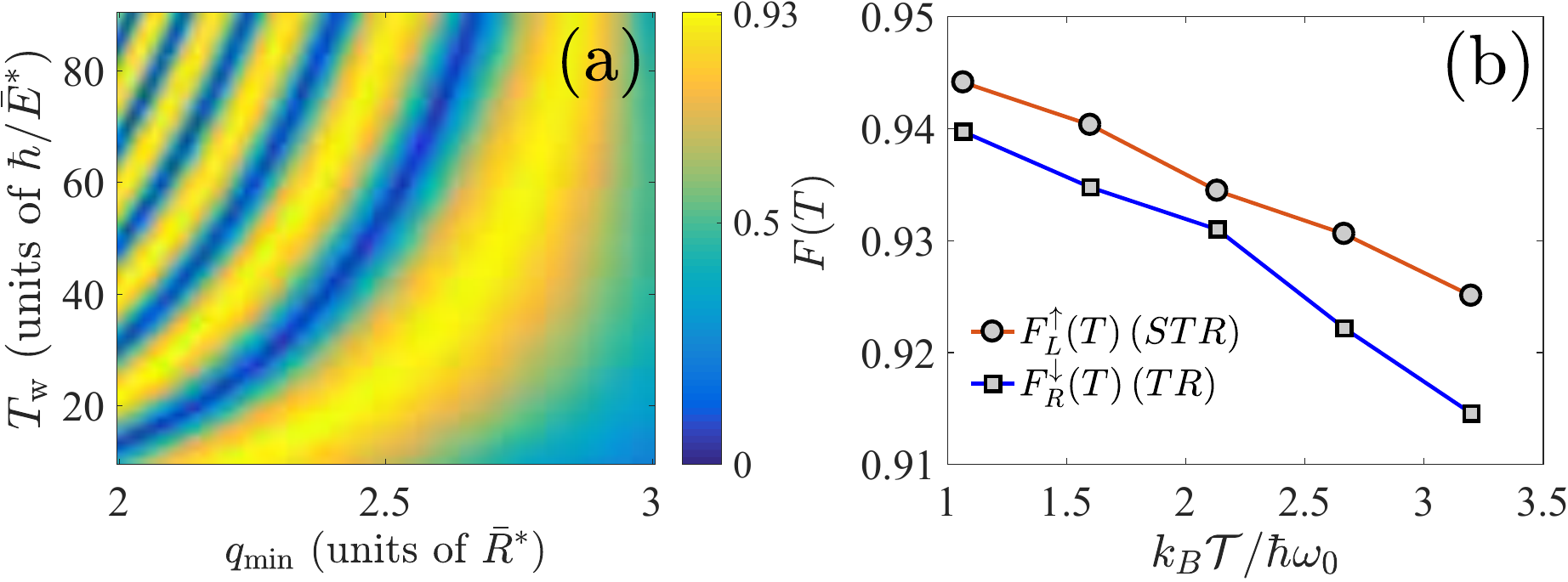}
\caption{(a) Uhlmann fidelity at the final time $T$ for a bosonic ensemble at the finite temperature $k_B\mathcal{T}=3.2\,\hbar\omega_0$ interacting with a static ion as a function of the waiting time $T_{\mathrm{w}}$ and minimal well separation from the barrier $q_{\min}$ for the tunneling regime. The barrier height is set to $b= 5.5 \bar{E}^*$. 
(b) Uhlmann fidelity at the final time $T$ for the TR and STR processes as a function of the quantum gas temperature $\mathcal{T}$ in units of the well quantum $\hbar\omega_0$. The trap parameters $q_{\min}$, $T_{\mathrm{w}}$ and $T_{\mathrm{r}}$ have been chosen as in Fig.~\ref{fig:f9}. The line connecting the points is merely a guide to the eye. In both panels we have $N=50$ atoms with $g=0.004$. The results are averaged over $N_s = 100$ stochastic fields. As a rule of thumb, the number of noncondensate atoms is about 13 for $k_B\mathcal{T} = 3.2\,\hbar\omega_0$.
}
\label{fig:finiteTstatic}
\end{figure}

The result of this study is summarized in Fig.~\ref{fig:finiteTstatic}, where in the left panel we show the Uhlmann fidelity for the tunneling process at $k_B\mathcal{T}=3.2\,\hbar\omega_0$ as a function of the separation and waiting time, while in the right panel the Uhlmann fidelity as a function of the temperature $\mathcal T$ is shown. First, we see that the fidelity shows also at finite temperature the same dependence on $q_{\min}$ and $T_{\mathrm{w}}$, as in the left of Fig.~\ref{fig:f7}. Notwithstanding, the maximum of the fidelity is no longer unity, but about 0.93. Hence, this indicates a strong impact of the gas temperature on the tunneling dynamics. Second, the temperature affects almost equally the TR and the STR (i.e., there is almost the same slope in the right panel), even though the fidelity of the TR is worse than that of the STR, as atoms do cross the barrier, and therefore do interact with the ion more strongly than in the STR. 
In contrast to imperfect ion ground state cooling (see Fig.~9 in Ref.~\cite{JogerPRA14}), however, the entanglement scheme is more resilient to thermal excitations in the Bose gas, thus enabling one to obtain the desired state at finite temperature still reasonably well. 

We continue the analysis by including the motional degree of freedom of the ion. When the ion is moving, however, the quantum dynamics of the combined system does not simply obey the GPE, as we have seen in the previous subsection, and therefore the TW method has to be revisited. Initially, however, when the Bose gas and the ion do not interact, the stochastic ensemble of classical fields can be generated as in the standard approach, that is, by using the same prescription we used for the static ion limit, as outlined in Ref.~\cite{SinatraPRL01}. The initial ensemble of stochastic fields is then propagated according to Eq.~(\ref{eq:eq20}) instead of Eq.~(\ref{eq:GPE}), where the ion and condensate motion are correlated. We justify this approach by the fact that we have assumed for each ionic motional state a condensate-like wave function, i.e., a product state, as for the GPE. Thus, in the same spirit of the original formulation of the TW method for single species, we apply it to the compound atom-ion quantum system for (partially) uncorrelated~\footnote{Here by ``partially'' we refer to the fact that, while the bosonic state for each $n$ motional state of the ion is uncorrelated, for $n \ne n^\prime$ the bosonic single-particle states are correlated via Eqs.~(\ref{eq:eq19}-\ref{eq:eq20}). Thus, albeit mean-field theory applies within each $n$ mode, it is not true for $n \ne n^\prime$.}, but interacting, bosons. In this framework, the one-body density matrix can be computed by using Eq.~(\ref{eq:rBG1body}) as
\begin{align}
\label{eq:rBG1bodycoordinate}
\rho_{\mathrm{BG}} (z,z^\prime) &= \langle z\vert \hat \rho_{\mathrm{BG}}\vert z^\prime\rangle \nonumber\\
\phantom{=}&= \sum_{n} \langle\vert c_n \vert^2\rangle_W (\langle \varphi_{n}^*(z) \varphi_{n}(z^\prime)\rangle_W - n_q \delta_{z,z^\prime})
\end{align}
where $\ket{\varphi_{n}}\bra{\varphi_{n}}$ has been replaced by the term in parentheses in the last line of Eq.~(\ref{eq:rBG1bodycoordinate}) in coordinate space representation. This expression is a weighted average over the vibrational states of the ion with weight $\langle\vert c_n \vert^2\rangle_W$, which is the statistical average over the $N_s$ values of $\vert c_n \vert^2$ (for each $n$) as a consequence of the $N_s$ stochastic fields generated initially. Similarly to Eq.~(\ref{eq:rhoG}), $\langle \varphi_{n}^*(z) \varphi_{n}(z^\prime)\rangle_W$ represents the statistical average over the same $N_s$ stochastic fields. Note also that $\rho_{\mathrm{BG}} (z,z^\prime)$ is a $\mathcal N\times \mathcal  N$ Hermitian matrix with $\mathcal N$ being the number of grid points. 

Now, since Eq.~(\ref{eq:eq20}) determines the time evolution of the single-particle orbitals of the bosons by assuming them to be of $O(1)$ norm, contrary to the usual particle-number-conserving Bogoliubov approach of TW~\cite{SinatraPRL01,CockburnPRA11}, where the stochastic fields have $O(N)$ norm, we proceed as follows: 
\begin{itemize}
\item[(i)] We first generate $N_s$ stochastic fields of $O(N)$ norm as in the original TW method. 
\item[(ii)] We propagate the $N_s$ stochastic fields according to the modified Eqs.~(\ref{eq:eq19},\ref{eq:eq20}), where their $O(N)$ norms are taken into account (see appendix~\ref{sec:Correquations}).
\item[(iii)] We compute the resulting Uhlmann fidelity at the final time $T$.  
\end{itemize}
In this way we can consistently use the equations of motion Eq.~(\ref{eq:eq20}), which, together with Eq.~(\ref{eq:eq19}), preserve unitarity, as well as utilize the prescription of Ref.~\cite{SinatraPRL01} for generating the initial sample of stochastic fields for the whole bosonic matter-wave field. 

We have applied the above outlined `recipe' in order to investigate the impact of finite temperature on the Uhlmann fidelity for the case of a moving ion, as discussed previously. In particular, we have considered an ion trap frequency of $\omega = 81\,\omega_0$ in interaction with $N=50$ bosons with coupling strength $g = 0.004 E^* R^*$. We have analyzed the TR~\footnote{For the STR we obtain essentially unity Uhlmann fidelity.} and found that, for example, for $k_B\mathcal{T} = 1.6\,\hbar\omega_0$ it is about 0.94, i.e., very similar to the static ion case [cf. panel (b) of Fig.~\ref{fig:finiteTstatic}]. Furthermore, we find $\langle\vert c_0 \vert^2\rangle_W\simeq 0.97$ and $\langle\vert c_1 \vert^2\rangle_W\simeq 0.03$, meaning that the ion is almost in its ground state. Hence, the combination of both ion motion and finite temperature of the Bose gas does not render the performance of the entanglement protocol worse. Thus, this indicates that finite temperature is indeed the major source of imperfections and quite cold atomic ensembles are required for a successful implementation of the protocol.


\section{Entanglement generation with the ion motional state}
\label{sec:motionEnta}

In this section we investigate the possibility to entangle the Bose gas with the motional states of the ion. Particularly, we investigate the zero-temperature scenario. 
Similarly to the mean-field analysis, we look for a suitable entanglement process by fixing $T_{\mathrm{r}}=5\hbar/\bar{E}^*$ and $q_{0}=5\bar{R}^*$, but by varying the $T_{\mathrm{w}}$ within the range $5 - 10.8\,\hbar/\bar{E}^*$ and $q_{\min}$ within the range $2.2 - 2.8\,\bar{R}^*$. Moreover, we consider $N=10$ atomic bosons with an interaction strength $g=0.02\, \bar{E}^* \bar{R}^*$. In this way we have the same $g N/(\bar{E}^* \bar{R}^*)=0.2$ as in Sec.~\ref{sec:Tzero}, but at the same time we reduce the complexity of the numerical simulation of the equations of motion~(\ref{eq:eq23}). Furthermore, the ion trap frequency has been chosen as $\omega = 12.8\,\omega_0$, the barrier height has been chosen as $b = 5.5\,\bar{E}^*$, and the ion internal state has been chosen as $\ket{\downarrow}$, whereas initially the atomic ensemble is prepared in the left well.

\begin{figure}
\includegraphics[scale=0.5]{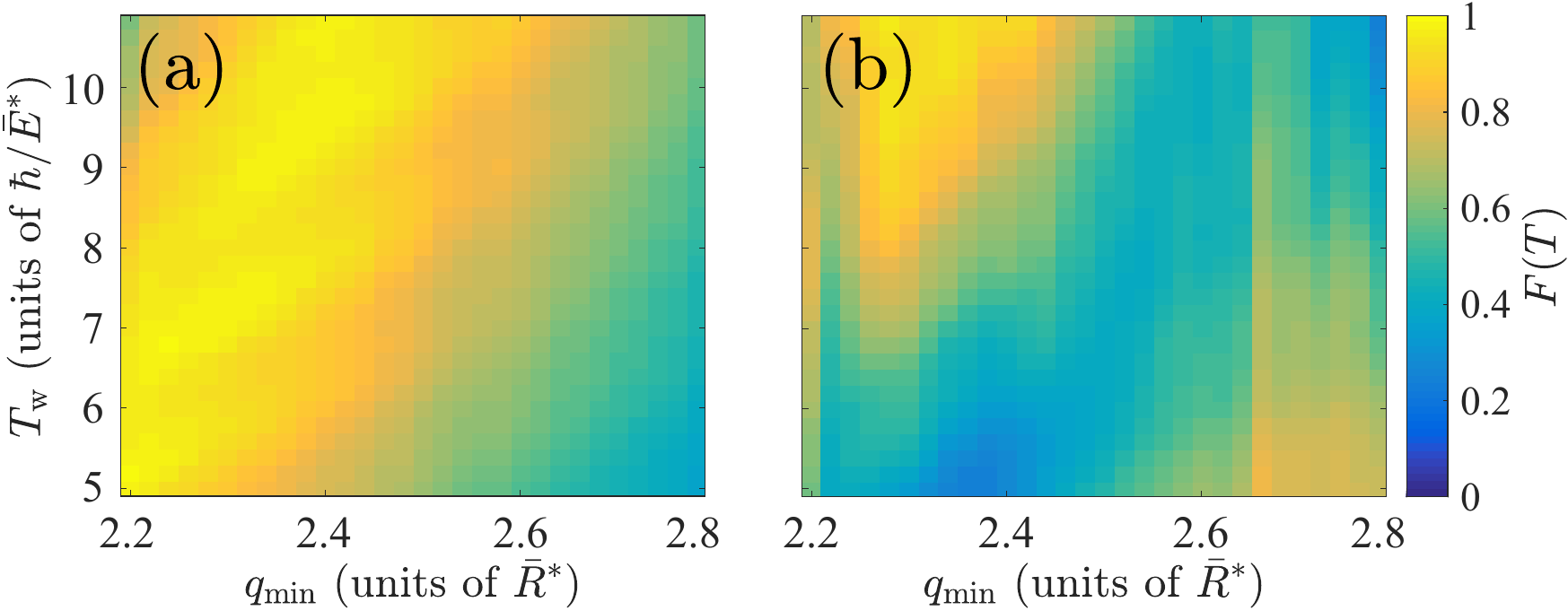}
\caption{\label{fig:f13} Uhlmann fidelity with the one-body density matrix of the bosons at the final time $T$ for the processes~(\ref{eq:gate}) with the ion motional states. In the panel (a), the fidelity $F_{R}(T)$ for the ion initially prepared in the ground state is shown, whereas in panel (b) we display the fidelity $F_L(T)$ for the ion initially prepared in the first excited state.}
\end{figure}

\begin{figure}
\includegraphics[scale=0.45]{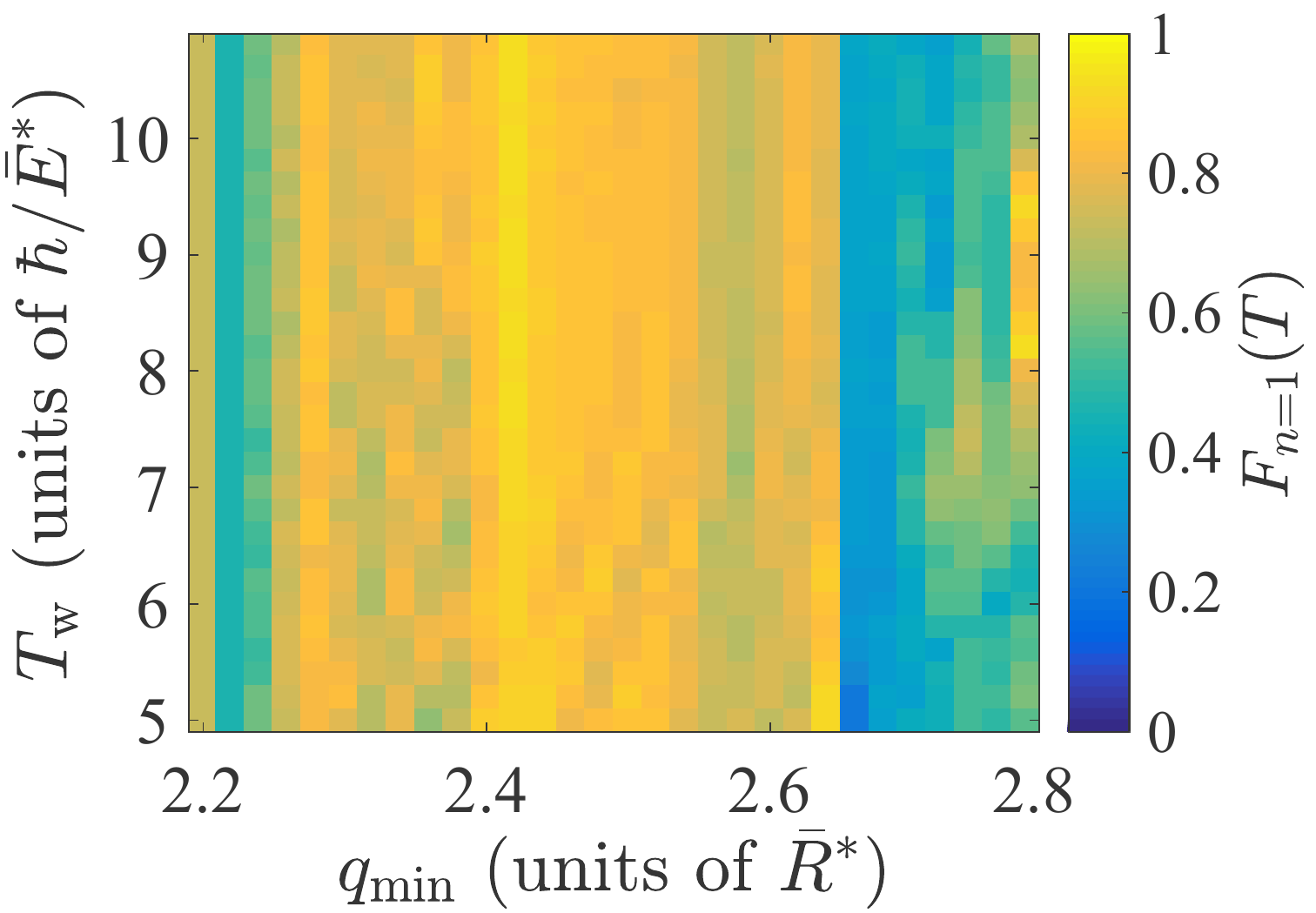}
\caption{\label{fig:f14} Uhlmann fidelity of the ion at the final time $T$, computed by using exactly the same formula~(\ref{eq:fidelity}), but with the ion-reduced density matrix defined in Eq.~(\ref{eq:rI}), for the entanglement protocol with the ion motional states. In particular, we display the fidelity for the ion initially prepared in the first excited state. For the ion initially prepared in the ground state, the fidelity is essentially unity within the same range of parameters.}
\end{figure}

In this scenario, we proceed in such a way that when the ion is initially prepared in the vibrational ground state ($n=0$) the one-body density matrix of the bosons is $\hat\rho_G =\ket{\psi_R}\bra{\psi_R}$, while when the ion is in the first excited state ($n=1$) of the trap the goal state is $\hat\rho_G =\ket{\psi_L}\bra{\psi_L}$. The results of this analysis are depicted in Figs.~\ref{fig:f13} and~\ref{fig:f14}. Interestingly, when the ion is in the ground state, then the ion motional state is almost unperturbed (the ion Uhlmann fidelity is generally larger than 0.98, not shown). On the other hand, when the ion is in the first excited state $(n=1)$, then the ion motion is substantially affected by the coupling to the bosonic ensemble. Similarly to the peaks we have observed in Fig.~\ref{fig:f12}, we attribute this difference to the rise of avoided crossings for excited vibrational states of the ion that cause admixture with excited motional states of both the ion and the bosons. Specifically, for the chosen trap frequencies, the set of parameters $T_{\mathrm{w}}=10.8\,\hbar/\bar{E}^*$ and $q_{\min}=2.42\,\bar{R}^*$ is `optimal', that is, $F_{R}(T)\simeq 0.97$ and $F_{L}(T)\simeq 0.92$ for the bosonic ensemble, while $F_{n=0}(T)\simeq 0.997$ and $F_{n=1}(T)\simeq 0.93$ for the ion. We have performed a similar analysis for $N=50$ and $g=0.004\, \bar{E}^* \bar{R}^*$ and found that it is still possible to identify suitable parameters for performing the transformations~(\ref{eq:gate}), but, particularly when the ion is prepared initially in the first excited state, the parameter range is narrower and precise control of the double-well position and waiting time is required. 
  
We conclude this section by underscoring that so far we have only investigated the performance of the two processes~(\ref{eq:gate}) by simulating the quantum many-body dynamics with the wave function ansatz~(\ref{eq:Psit}), once for the initial state $\ket{\Xi_{\mathrm{initial}}^0} \equiv \ket{\psi_L}\ket{n=0}$ and once for $\ket{\Xi_{\mathrm{initial}}^1} \equiv \ket{\psi_L}\ket{n=1}$. Of course, within the time window $0< t < T$ the resulting atom-ion quantum dynamics is highly correlated involving occupations of different motional ion states. Since, however, at the final time $T$ the resulting states are very close to either $\ket{\psi_R}\ket{n=0}$ (for $\ket{\Xi_{\mathrm{initial}}^0}$) or $\ket{\psi_L}\ket{n=1}$ (for $\ket{\Xi_{\mathrm{initial}}^1}$), because of the linearity of the Schr\"odinger equation we can be sure that an entangled atom-ion state very close to $c_0 \ket{\psi_R}\ket{n=0} + c_1 \ket{\psi_L}\ket{n=1}$ can be attained for specific values of $q_{\min}$ and $T_{\mathrm{w}}$ and for any initial superposition $c_0\ket{n=0}+c_1\ket{n=1}$ of the ion motional states. To corroborate further this fact, we have performed a numerical simulation with the initial state
\begin{align}
\ket{\Xi_{\mathrm{initial}}} = \frac{1}{\sqrt{2}} \prod_{j=1}^N\ket{\psi_L^{(j)}} (\ket{n=0} + \ket{n=1})
\end{align} 
where the state $\prod_{j=1}^N\ket{\psi_L^{(j)}}$ means that all bosons are in the same single-particle wave function $\psi_L(z_j)$, whereas the ion is in an equal superposition of the two lowest-energy states of the harmonic trap. On the other hand, the time evolved state is given by Eq.~(\ref{eq:Psit}) and the target state is the equal atom-ion superposition 
\begin{align}
\ket{\Xi_{\mathrm{target}}} = \frac{1}{\sqrt{2}} \left(\prod_{j=1}^N\ket{\psi_R^{(j)}} \ket{n=0} + \prod_{j=1}^N\ket{\psi_L^{(j)}} \ket{n=1}\right).
\end{align} 
In order to quantify the quality of the state at the final time $T$, which is obtained by means of the optimal double-well separation $q(t)$ we found previously for the two single processes, we evaluate the Uhlmann fidelities of the bosons and ion, respectively. This means that for both the resulting state from the dynamics at time $T$ and the target state $\ket{\Xi_{\mathrm{target}}}$ given above, we assess the corresponding one-body density matrices and compare them. Specifically, we obtain $F_{R}(T)\simeq 0.95$ and $F_{L}(T)\simeq 0.93$ for the bosonic ensemble and $F_{n=0}(T)\simeq 0.60$ and $F_{n=1}(T)\simeq 0.39$ for the ion [ideally $F_{n=0,1}(T)$ should be 0.5], which yield for the superposition state the fidelity estimate: $F_{n=0}(T)F_{R}(T)+F_{n=1}(T)F_{L}(T)\simeq 0.93$. 
In addition to this, we performed a finite-temperature simulation at $k_B\mathcal{T} = 1.6\,\hbar\omega_0$ and obtained the following Uhlmann fidelities for the bosonic ensemble (averaged over 100 stochastic fields): $F_{R}(T)\simeq 0.93$, $F_{L}(T)\simeq 0.91$. On the other hand, for the ion it is essentially the same as at zero temperature. Thus, the superposition state fidelity becomes approximatively 0.91. 
These numbers confirm that the desired entangled state is attained sufficiently well, i.e., at the level of the one-body density, and that it is indeed sufficient to optimize the two processes separately. Moreover, the thermal fluctuations of the bosonic ensemble affect only its dynamics, namely the ion motional state is unaffected, at least at low temperatures. To obtain better performances, however, additional shaping of the transport function $q(t)$ can be attained by means of optimal control techniques~\cite{Negretti2013,BrouzosPRA15}. Those methods allow not only for enhanced performance of the entanglement protocol, but also for increased robustness against experimental imperfections of the manipulated time-dependent functions~\cite{NegrettiJPB11}. 

\section{Conclusions and outlook}
\label{sec:outlook}

We have investigated the quantum dynamics of an atomic degenerate Bose gas in a double-well potential coupled to a trapped ion. Contrary to previous studies concerning the same setup~\cite{GerritsmaPRL12,SchurerPRA16}, we have analyzed in detail the impact of the ion motion on the dynamics of the ultracold quantum gas as well as the impact of finite temperature of the gas on the entanglement protocol. We have found that, upon the tightness of the ion trap, different vibrational states of the ion motion can be populated. This is particularly true for the tunneling regime, as the atoms interact more frequently with the ion as they pass through the barrier, whereas for the self-trapping regime, even shallower ion traps do not affect significantly the bosonic dynamics in the well. We have also found that thermal excitations on the top of the condensate can have a detrimental effect on the protocol. Notwithstanding, a suitable entanglement scheme can be attained, if we do not aim at populating a single mode of the bosonic field (i.e., the condensate mode) as well as if the ion is not strictly kept in the ground state. Indeed, entanglement between two quantum systems can be also attained for mixed states~\cite{HorodeckiRMP09}. In addition to this, we have demonstrated the possibility to realize an entanglement protocol between the ion and the bosonic ensemble motions, and therefore beyond the two-body schemes proposed in previous studies~\cite{JogerPRA14,DoerkPRA10}. This alternative pathway to entangle many-particle systems with the quantized motion of an ion is particularly appealing. Indeed, recent experiments~\cite{RatschbacherPRL13,FuerstPRA18,SikorskyNatCom18,SikorskyPRL18} on the ion spin dynamics have shown that ionic spin relaxation, which is caused by large second-order spin-orbit interactions~\cite{TscherbulPRL15}, occurs after a few Langevin collisions. Hence, this is indeed a fundamental drawback for the recently proposed entanglement schemes based on ionic spin-dependent short-range interactions. On the other hand, the experimentally observed strong spin-exchange processes~\cite{FuerstPRA18} could be exploited to develop alternative entanglement schemes. 

Another interesting aspect of our paper is the application of the truncated Wigner method for simulating a Bose gas at finite temperature in the presence of an impurity. We have applied the method $ad$ $hoc$ and made quantitative predictions on the fidelity of the two processes~(\ref{eq:gate}). In the future, it would be definitely interesting to better understand the applicability and validity of our strategy to describe the dynamics at nonzero temperature of systems involving an impurity in a thermal bath. 
This would have relevant applications in current research on impurity physics with atomic quantum gases, e.g., in order to better understand the role of finite-temperature effects on the polaron formation, especially in view of its coherence properties, which are typically analyzed in the framework of the (Markovian) quantum master equation formalism~\cite{KrychPRA15,Nielsen2018}. 

Finally, a natural extension of the present paper concerns the replacement of the quantum Bose gas with an ensemble of spin-polarized fermionic atoms, since experiments combining trapped ytterbium ions with fermionic lithium atoms are becoming available~\cite{FuerstPRA18,JogerPRA17}. Apart from the generation of entangled atom-ion states, an interesting question in this regard is how the ion can modify the junction transport properties. For example, can an ion stop the atomic flow through the barrier, particularly when the system is a superfluid, or can $\mu$m-sized density bubbles in the Fermi gas be formed, similarly to an ion embedded in a strongly correlated Tonks-Girardeau atomic gas~\cite{GooldPRA10}? Besides this, another intriguing research direction is the generation of correlated quantum dynamics induced by time-dependent modulation of the atom-ion interaction via laser fields that couples to Rydberg states~\cite{SeckerPRA16,SeckerPRL17}, as conducted in recent pioneering experiments~\cite{Ewald2018,Haze2019}. Here, interesting questions on thermalization of closed systems and energy exchange between subsystems could be especially addressed. 


\section{Acknowledgements}

We are very grateful to Mrs. F. Jafariani for helping us with the design of the figures. We are also very thankful to Angela White for her critical reading of the paper and suggestions. M.~R.~E. acknowledges the Zentrum f\"ur Optische Quantentechnologien at the Universit\"at Hamburg, where this project has been conceived, for the kind hospitality, while A. N. acknowledges R. Gerritsma, J. M. Schurer, F. Schmidt-Kaler, Z. Idziaszek, and T. Calarco for collaboration on related previous works. Finally, this work was supported by the Cluster of Excellence projects "The Hamburg Centre for Ultrafast Imaging" of the Deutsche Forschungsgemeinschaft (EXC 1074, Project No. 194651731) and "CUI: Advanced Imaging of Matter" of the Deutsche Forschungsgemeinschaft (EXC 2056, Project No. 390715994), by the Ministry of Science Research and Technology of Iran, and by the Iran Science Elites Federation. 


\appendix 
\section{Appendixes}
\label{sec:appendix}

In this appendix we derive the equations of motion for a Bose gas in an external trap (e.g., a double well) interacting with an ion confined in a harmonic trap, as we discussed in Sec.~\ref{sec:Tzero}. In doing so, we shall provide all steps of the derivation in a pedagogical manner, given the fact that the approach based on the Dirac-Frenkel variational principle that we adopt here might not be so familiar and that the technicalities of the derivation of the corresponding equations of motion are usually skipped. In addition to this, we also discuss briefly the actual numerical implementation. 


\subsection{Hamiltonian}

Let us consider the one-dimensional Hamiltonian describing $N$ interacting bosons of mass $m$ in interaction with an impurity of mass $M$ such as an ion:
\begin{align}
\hat H & =  \sum_{j=1}^N \left[
\frac{\hat p^2_j}{2 m} + V_{\mathrm{ext}}(z_j) + V_{\mathrm{ai}}(z_j-Z)
\right] \nonumber\\
\phantom{=} & + g\sum_{j < i}\delta(z_j - z_i) + \frac{\hat P^2}{2 M} + \frac{M\omega^2}{2}  Z^2.
\end{align}
Here $\hat p_j=-i\hbar\partial_{z_j}$ is the momentum operator of the $j$-th boson, $V_{\mathrm{ext}}(z_j)$ is an external confining potential for the bosons, 
$V_{\mathrm{ai}}(z_j-Z)$ describes the interaction between the bosons and the impurity atom, $\hat P=-i\hbar\partial_Z$ is the impurity's momentum operator, 
and $\omega$ is the impurity trap frequency. Here $\partial_Z$ denotes the first partial derivative with respect to $Z$.

The Hamiltonian rescaled with respect to the length $\bar{R}^{*}$ and energy $\bar{E}^{*}$ reads
\begin{equation}
\label{eq:eq2}
\begin{aligned}
{\bar{H}} =& \sum_{j=1}^{N}\left[-\dfrac{\partial^{2}}{\partial \bar{z}_j^{2}}+\bar{V}_{\mathrm{ext}}(\bar{z}_j)+ \bar{V}_{\mathrm{ai}}(\bar{z}_j-\bar{Z})\right]\\
& +\dfrac{\overline{g}}{2}\sum_{k\neq j}\delta (\bar{z}_j - \bar{z}_k) + {\bar{H}}_{\mathrm{i}},
\end{aligned}
\end{equation}
with $\bar{z}_j = z_j/\bar{R}^*$, $\bar{V}_{\mathrm{ext}}(\bar{z}_j) = V_{\mathrm{ext}}(\bar{z}_j)/\bar{E}^*$, etc., and
\begin{align}
\bar{H}_{\mathrm{i}}=-\dfrac{m}{M}\dfrac{\partial^{2}}{\partial \bar{Z}^{2}} + \dfrac{m}{M}\bigg(\dfrac{\bar{R}^{*}}{\ell_{i}}\bigg)^{4}\bar{Z}^{2},
\end{align}
where $\ell_{i}=[\hbar/(M\omega)]^{1/2}$, and $\bar{H}_{\mathrm{i}}\vert\phi_{n}\rangle=\bar{E}_{n}\vert\phi_{n}\rangle$. 


\subsection{Many-body state ansatz}

The general many-body quantum state for such a compound system, which takes into account all correlations, is given by 
\begin{equation}
\label{eq:eq3}
\vert\Psi\rangle=\sum_{n=0}^{\infty} c_{n}\vert\Phi_{n}\rangle\vert\phi_{n}\rangle,
\end{equation}
with $\sum_{n=0}^{\infty}\vert c_{n}\vert^{2}=1$. In practice we truncate the impurity Hilbert space in such a way that only the first $n_{i}+1$ states are relevant. In order to proceed further, we make the assumption that the bosons are accurately described by the tensor product state $\vert\Phi_{n}\rangle=\prod_{j=1}^{N}\vert\varphi_{n}^{(j)}\rangle$. We would like to underscore that while within such kind of mean-field approximation we neglect correlations between bosons when the impurity is in the $n$-th motional state, bosonic correlations can occur when the ion is occupying simultaneously different motional states. Under these conditions, the total system state is given by
\begin{equation}\label{eq:eq4}
\vert\Psi\rangle=\sum_{n=0}^{n_{i}}c_{n}\prod_{j=1}^{N}\vert\varphi_{n}^{(j)}\rangle\vert\phi_{n}\rangle,
\end{equation}
with $\langle\varphi_{n}\vert\varphi_{n}\rangle=1  \hspace{0.2cm} \forall  n$. We note, however, that the single-particle states (i.e., orbitals) $\vert\varphi_{n}\rangle$ are in general nonorthogonal, while we consider an orthonormal and time-independent basis for the impurity states $\vert\phi_{n}\rangle$. In this formulation, the coefficients $c_n$ and the orbitals $\vert\varphi_{n}\rangle$ are the only time-dependent functions. 


\subsection{Derivation of the equations of motion}
\label{sec:Correquations}

In order to obtain differential equations for $c_n$ and $\vert\varphi_{n}\rangle$ we employ the so-called Dirac-Frenkel variational principle~\cite{DiracMPCPS30,Frenkel34}
\begin{equation}
\label{eq:eq5}
\langle\delta\Psi\vert i \partial_{\tau}-\bar{H}\vert\Psi\rangle=0,
\end{equation}
with $\tau=(\bar E^*/\hbar)\,t$. Here, $\delta\Psi$ denotes the variation of the total wave function with respect to the free parameters $c_{n}$ and $\varphi_{n}$, while we ignore any variation of the ion's states $\phi_{n}$, as we have chosen them to be time independent. 



Towards that end, we begin with the computation of the vector state $\langle\delta\Psi\vert$, which is simply given by
\begin{equation}\label{eq:eq6}
\langle\delta\Psi\vert\!=\!\sum_{n=0}^{n_{i}}\!\!\left[\delta c_{n}^{*}\prod_{j=1}^{N}\langle \varphi_{n}^{(j)}\vert \langle\phi_{n}\vert + c_{n}^{*}\sum_{j=1}^{N}\prod_{\substack{i=1 \\i\neq j}}^{N}\langle \delta\varphi_{n}^{(j)}\vert \langle \varphi_{n}^{(i)}\vert \langle\phi_{n}\vert\right]
\end{equation}
whereas for the time derivative of the state we obtain
\begin{equation}\label{eq:eq7}
\vert\dot\Psi\rangle=\sum_{n=0}^{n_{i}}\dot c_{n}\prod_{j=1}^{N}\vert \varphi_{n}^{(j)}\rangle \vert\phi_{n}\rangle + \sum_{n=0}^{n_{i}} c_{n}\sum_{j=1}^{N}\prod_{\substack{i=1 \\i\neq j}}^{N}\vert \dot\varphi_{n}^{(j)}\rangle \vert \varphi_{n}^{(i)}\rangle \vert\phi_{n}\rangle.
\end{equation}

Now, let us assess the two scalar products involved in Eq. (\ref{eq:eq5}). To this aim, we start with the scalar product between the above outlined state variation and the time derivative of the many-body state, which yields
\begin{eqnarray}\label{eq:eq8}
\langle\delta\Psi\vert\dot\Psi\rangle=&& \sum_{n=0}^{n_{i}}\delta c_{n}^{*}\Big[\dot c_{n}+N c_{n}\langle\varphi_{n}\vert\dot \varphi_{n}\rangle\Big]\nonumber\\
&& +N\sum_{n=0}^{n_{i}}\langle\delta\varphi_{n}\vert \Big\{ c_{n}^{*}\dot c_{n}\vert\varphi_{n}\rangle + \vert c_{n}\vert^{2} \nonumber\\
&&\times\Big[ \vert\dot\varphi_{n}\rangle +
(N-1)\langle\varphi_{n}\vert\dot\varphi_{n}\rangle\vert\varphi_{n}\rangle\Big]\Big\}.\nonumber\\
\end{eqnarray}
We then compute the second scalar product in Eq. (\ref{eq:eq5}), which is a sort of expectation value of the many-body Hamiltonian that gives the following result:
\begin{widetext}
\begin{align}
\label{eq:eq9}
&\langle\delta\Psi\vert \bar{H}\vert\Psi\rangle= \sum_{n=0}^{n_{i}}\delta c_{n}^{*}\Big[c_{n}N\langle\varphi_{n}\vert H_{0}\vert\varphi_{n}\rangle 
+\dfrac{\overline{g}}{2}N(N-1)c_{n}\langle\varphi_{n},\varphi_{n}\vert \delta(\bar{z}-\bar{y})\vert\varphi_{n},\varphi_{n}\rangle +c_{n}\bar{E}_{n}
\nonumber\\
& +\sum_{n^{\prime}=0}^{n_{i}}c_{n^{\prime}}N\langle\varphi_{n},\phi_{n}\vert \bar{V}_{\mathrm{ai}}\vert\phi_{n^{\prime}},\varphi_{n^{\prime}}\rangle
 \langle\varphi_{n}\vert\varphi_{n^{\prime}}\rangle^{N-1}
 \Big]
+N \sum_{n=0}^{n_{i}}\vert c_{n}\vert^{2} \langle\delta\varphi_{n}\vert H_{0}\vert\varphi_{n}\rangle
+\sum_{n=0}^{n_{i}}\langle\delta\varphi_{n}\vert\Big[\vert c_{n}\vert^{2}N(N-1)\langle\varphi_{n}\vert H_{0}\vert\varphi_{n}\rangle\Big]\vert\varphi_{n}\rangle\nonumber\\
& +\dfrac{\overline{g}}{2} N \sum_{n=0}^{n_{i}}\vert c_{n}\vert^{2}\Big[2(N-1)\langle\varphi_{n}\vert\delta(\bar{z}-\bar{y})\vert\varphi_{n}\rangle  +(N^{2}-3N+2)\langle\varphi_{n},\varphi_{n}\vert\delta(\bar{z}-\bar{y})\vert\varphi_{n},\varphi_{n}\rangle\Big] \langle\delta\varphi_{n}\vert\varphi_{n}\rangle  
+N \sum_{n=0}^{n_{i}}\vert c_{n}\vert^{2}\bar{E}_{n} \langle\delta\varphi_{n}\vert\varphi_{n}\rangle
\nonumber\\&  
+N \sum_{n,n^{\prime}=0}^{n_{i}} c_{n}^{*}c_{n^{\prime}}
\langle\delta\varphi_{n},\phi_{n}\vert \bar{V}_{\mathrm{ai}}\vert\phi_{n^{\prime}},\varphi_{n^{\prime}}\rangle\langle\varphi_{n}\vert\varphi_{n^{\prime}}\rangle^{N-1}
+N(N-1)\sum_{n,n^{\prime}=0}^{n_{i}}c_{n}^{*}c_{n^{\prime}}
\langle\varphi_{n},\phi_{n}\vert \bar{V}_{\mathrm{ai}}\vert\phi_{n^{\prime}},\varphi_{n^{\prime}}\rangle\langle\varphi_{n}\vert
\varphi_{n^{\prime}}\rangle^{N-2}
\langle\delta\varphi_{n}\vert \varphi_{n^{\prime}}\rangle,\nonumber\\
\end{align}
\end{widetext}
with $ H_{0}=-\partial^2_{\bar{z}}+\bar{V}_{\mathrm{ext}}$. Let us note the appearance of the factor 
$\langle\varphi_{n}\vert\varphi_{n^{\prime}}\rangle^{N-k}$ with $k=1,2$. This is a direct consequence of the product ansatz~(\ref{eq:eq4}) we made for the ensemble of bosons. Upon the ``strength" of the orthogonality between the orbitals $\varphi_{n}(\bar{z})$ and $\varphi_{n^{\prime}}(\bar{z})$ and the number of bosons, terms in Eq. (\ref{eq:eq9}) in which such overlaps appear will be large or small, thus strongly influencing the correlated impurity-gas quantum dynamics. 

Now that we have all ingredients, we can insert the results of Eqs. (\ref{eq:eq8}) and (\ref{eq:eq9}) into Eq. (\ref{eq:eq5})
in order to obtain the following equations of motion for the expansion coefficients

\begin{align}
\label{eq:eq10}
& i\Big[\dot c_{n} +Nc_{n}\langle\varphi_{n}\vert\dot \varphi_{n}\rangle\Big]=
c_{n}(N\langle\varphi_{n}\vert H_{0}\vert\varphi_{n}\rangle + \bar{E}_{n})\nonumber\\
& +\dfrac{\overline{g}}{2}N(N-1)c_{n}\langle\varphi_{n},\varphi_{n}\vert\delta(\bar{z}-\bar{y})\vert\varphi_{n},\varphi_{n}\rangle\nonumber\\
& +\sum_{n^{\prime}=0}^{n_{i}}c_{n^{\prime}}N
\langle\varphi_{n},\phi_{n}\vert \bar{V}_{\mathrm{ai}}\vert\phi_{n^{\prime}},\varphi_{n^{\prime}}\rangle\langle\varphi_{n}\vert
\varphi_{n^{\prime}}\rangle^{N-1},
\end{align}
whereas for the single-particle wavefunctions of the bosons we obtain
\begin{widetext}
\begin{eqnarray}
\label{eq:eq11}
&& i\Big\{c_{n}^{*}\dot c_{n}\vert\varphi_{n}\rangle+\vert c_{n}\vert^{2}\Big[\vert\dot \varphi_{n}\rangle + (N-1)\langle\varphi_{n}\vert\dot \varphi_{n}\rangle\vert\varphi_{n}\rangle
\Big]\Big\}=\nonumber\\
&& \vert c_{n}\vert^{2}\Big\{H_{0} + (N-1)\langle\varphi_{n}\vert H_{0}\vert\varphi_{n}\rangle
 + \dfrac{\overline{g}}{2}\Big[2(N-1)\langle\varphi_{n}\vert\delta
(\bar{z}-\bar{y})\vert\varphi_{n}\rangle + (N^{2}-3N+2)\langle\varphi_{n},\varphi_{n}\vert\delta(\bar{z}-\bar{y})\vert
\varphi_{n},\varphi_{n}\rangle\Big] + \bar{E}_{n}\Big\}\vert\varphi_{n}\rangle\nonumber\\
&& +\sum_{n^{\prime}=0}^{n_{i}}c_{n}^{*}c_{n^{\prime}}\Big\{\langle\phi_{n}\vert \bar{V}_{\mathrm{ai}}\vert\phi_{n^{\prime}}\rangle\langle\varphi_{n}\vert\varphi_{n^{\prime}}\rangle^{N-1}
+(N-1)\langle\varphi_{n},\phi_{n}\vert \bar{V}_{\mathrm{ai}}\vert\phi_{n^{\prime}},\varphi_{n^{\prime}}\rangle\langle\varphi_{n}
\vert\varphi_{n^{\prime}}\rangle^{N-2}\Big\}\vert\varphi_{n^{\prime}}\rangle.
\end{eqnarray} 
\end{widetext}
In order to arrive at the above outlined equations, we note that we made use of the following identities
\begin{equation}
\label{eq:eq12}
\begin{aligned}
\langle\varphi_{n}\vert\delta(\bar{z}-\bar{y})\vert\varphi_{n}\rangle=\vert\varphi_{n}(\bar{z})\vert^{2},
\end{aligned}
\end{equation} 
\begin{equation}
\label{eq:eq13}
\begin{aligned}
\langle\varphi_{n},\varphi_{n}\vert\delta(\bar{z}-\bar{y})\vert\varphi_{n},\varphi_{n}\rangle=\int d\bar{z} \vert\varphi_{n}(\bar{z})\vert^{4}.
\end{aligned}
\end{equation} 
Now, by substituting in Eq.~(\ref{eq:eq11}) the time derivative $\dot c_{n}$ obtained in Eq.~(\ref{eq:eq10}), we finally arrive at 
\begin{equation}
\label{eq:eq14}
\begin{aligned}
i\vert\dot \varphi_{n}\rangle=& i\langle\varphi_{n}\vert\dot\varphi_{n}\rangle\vert\varphi_{n}\rangle
+H_{gp}^{n}[\varphi_{n}]\vert\varphi_{n}\rangle-\langle H_{gp}^{n}[\varphi_{n}]\rangle\vert\varphi_{n}
\rangle\\
& +\sum_{n=0}^{n_{i}}\dfrac{c_{n}^{*}c_{n^{\prime}}}{\vert c_{n}\vert^{2}}\Big\{\langle\varphi_{n},\phi_{n}\vert \bar{V}_{\mathrm{ai}}\vert\phi_{n^{\prime}},\varphi_{n^{\prime}}\rangle(N-1)\vert
\varphi_{n^{\prime}}\rangle\\
& -N\langle\varphi_{n},\phi_{n}\vert \bar{V}_{\mathrm{ai}}\vert\phi_{n^{\prime}},\varphi_{n^{\prime}}\rangle\langle\varphi_{n}\vert\varphi_{n^{\prime}}\rangle\vert\varphi_{n}\rangle\\
& +\langle\phi_{n}\vert \bar{V}_{\mathrm{ai}}\vert\phi_{n^{\prime}}\rangle\langle\varphi_{n}\vert
\varphi_{n^{\prime}}\rangle\vert\varphi_{n^{\prime}}\rangle\Big\}\langle
\varphi_{n^{\prime}}\vert\varphi_{n^{\prime}}\rangle^{N-2},\\
\end{aligned}
\end{equation}
where $H_{gp}^{n}[\varphi_{n}]=-\dfrac{\partial^{2}}{\partial \bar{z}^{2}} + \bar{V}_{\mathrm{ext}}+\overline{g}(N-1)\vert\varphi_{n}\vert^{2}$ and $\langle H_{gp}^{n}[\varphi_{n}]\rangle=\langle\varphi_{n}\vert H_{gp}^{n}[\varphi_{n}]\vert\varphi_{n}\rangle$. With these definitions, we can rewrite Eq.~(\ref{eq:eq10}) as
\begin{equation}
\label{eq:eq15}
\begin{aligned}
i\dot c_{n}=& N c_{n}(\langle H_{gp}^{n}[\varphi_{n}/\sqrt{2}]\rangle -i\langle\varphi_{n}\vert\dot\varphi_{n}\rangle)\\
& \sum_{n^{\prime}=0}^{n_{i}}c_{n^{\prime}}N\langle\varphi_{n},\phi_{n}\vert
\bar{V}_{\mathrm{ai}}\vert\phi_{n^{\prime}},\varphi_{n^{\prime}}\rangle\langle\varphi_{n}\vert
\varphi_{n^{\prime}}\rangle^{N-1} \\
&+ c_{n}\bar{E}_{n},
\end{aligned}
\end{equation}
where $H_{gp}^{n}[\varphi_{n}/\sqrt(2)]=H_{0}+\dfrac{\overline{g}}{2}(N-1)\vert\varphi_{n}\vert^{2}$.
To further simplify the above outlined equations of motion, we first perform a unitary transformation on the coefficients, that is, $c=\hat U C$ with $c\cong(c_{0},c_{1},...,c_{n_{i}})^{T}$ 
(similarly for $C$), where 
\begin{equation}
\label{eq:eq16}
\hat U=
\begin{bmatrix}
 e^{-i\eta_{0}(t)} & 0 & . &\hspace{0.5cm} . & .\\
0 & e^{-i\eta_{1}(t)} & . &\hspace{0.5cm} . & . \\
. & . & . &\hspace{0.5cm} . & . \\
. & . & . &\hspace{0.5cm} . & . \\
. & . & . &\hspace{0.5cm} . & e^{-i\eta_{n_{i}}(t)}\\
\end{bmatrix}.
\end{equation}
By applying this unitary transformation we obtain
\begin{equation}
\label{eq:eq17}
\begin{aligned}
i\dot C_{n}=
&\sum_{n^{\prime}=0}^{n_{i}}C_{n^{\prime}}e^{-i(\eta_{n^{\prime}}(t)-\eta_{n}(t))}N
\langle\varphi_{n},\phi_{n}\vert \bar{V}_{\mathrm{ai}}\vert\phi_{n^{\prime}},\varphi_{n^{\prime}}\rangle \\
&\times\langle\varphi_{n}\vert\varphi_{n^{\prime}}\rangle^{N-1},
\end{aligned}
\end{equation}
\begin{equation}
\label{eq:eq18}
\begin{aligned}
\eta_{n}(t)=& N\int_{0}^{t}d\tau (\langle H_{gp}^{n}[\varphi_{n}/\sqrt{2}]\rangle-i\langle\varphi_{n}\vert\dot\varphi_{n}\rangle + \bar{E}_{n}/N)\\
& =N f_{n}(t).
\end{aligned}
\end{equation}
By defining $\vert\psi_{n}\rangle:=\vert\varphi_{n}e^{-if_{n}(t)}\rangle$, we have
\begin{equation}
\label{eq:eq19}
i\dot C_{n}=\sum_{n^{\prime}=0}^{n_{i}}C_{n^{\prime}}N\langle\psi_{n},\phi_{n}\vert \bar{V}_{\mathrm{ai}}\vert\phi_{n^{\prime}},\psi_{n^{\prime}}\rangle
\langle\psi_{n}\vert\psi_{n^{\prime}}\rangle^{N-1}.
\end{equation}
Since $\vert\varphi_{n}\rangle=e^{if_{n}(t)}\vert\psi_{n}\rangle$, we can use Eq.~(\ref{eq:eq14}) in order to get a differential equation for the quantum state $\vert\psi_{n}\rangle$, which is given by the following expression 
\begin{widetext}
\begin{eqnarray}
\label{eq:eq20}
i\vert\dot\psi_{n}\rangle=&& H_{gp}^{n}[\psi_{n}]\vert\psi_{n}\rangle
 + \sum_{n^{\prime}=0}^{n_{i}}\dfrac{C_{n}^{*}C_{n^{\prime}}}{\vert C_{n}\vert^{2}}\Big\{(N-1)\langle\psi_{n},\phi_{n}\vert \bar{V}_{\mathrm{ai}}\vert\phi_{n^{\prime}},\psi_{n^{\prime}}\rangle\langle
\psi_{n}\vert\psi_{n^{\prime}}\rangle^{N-2} 
+ \langle\phi_{n}\vert \bar{V}_{\mathrm{ai}}\vert\phi_{n^{\prime}}\rangle\langle\psi_{n}\vert\psi_{n^{\prime}}\rangle^{N-1} \Big\}
\vert\psi_{n^{\prime}}\rangle\nonumber\\
&& -\left\{\sum_{n^{\prime}=0}^{n_{i}}\dfrac{C_{n}^{*}C_{n^{\prime}}}{\vert C_{n}\vert^{2}}\Big[N
\langle\psi_{n},\phi_{n}\vert \bar{V}_{\mathrm{ai}}\vert\phi_{n^{\prime}},\psi_{n^{\prime}}\rangle\langle\psi_{n}\vert\psi_{n^{\prime}}\rangle^{N-1}\Big]
+\dfrac{\overline{g}}{2}(N-1)\langle\vert\psi_{n}\vert^{2}\rangle - \dfrac{\bar{E}_{n}}{N}\right\}\vert\psi_{n}\rangle.
\end{eqnarray}
\end{widetext}
In summary, we have derived coupled differential equations for the coefficients~(\ref{eq:eq19}) and for the orbitals~(\ref{eq:eq20}) that have to be solved numerically for given initial conditions. These equations, within the accuracy of the product ansatz for the bosons we made at the beginning, which is well justified for the purposes of our paper, provide a sufficiently good description of the compound quantum system dynamics. 

Finally, we note that Eqs.~(\ref{eq:eq19},\ref{eq:eq20}) are valid for zero temperature and for functions $\psi_{n}(x)$ normalized to unity. These equations, however, can be also utilized for the finite-temperature TW method as described in Sec.~\ref{sec:temperature}, but the following replacements are required:
\begin{align}
\langle\psi_{n}\vert\psi_{n^{\prime}}\rangle^{N-1}\mapsto \frac{\langle\psi_{n}\vert\psi_{n^{\prime}}\rangle^{N-1}}{\langle\psi_{n}\vert\psi_{n}\rangle^{N}}
\end{align}
in Eq.~(\ref{eq:eq19}), while in Eq.~(\ref{eq:eq20})
\begin{align}
\langle\psi_{n}\vert\psi_{n^{\prime}}\rangle^{N-k}\mapsto \frac{\langle\psi_{n}\vert\psi_{n^{\prime}}\rangle^{N-k}}{\langle\psi_{n}\vert\psi_{n}\rangle^{N-k+1}}
\end{align}
for the terms in which $\langle\psi_{n},\phi_{n}\vert \bar{V}_{\mathrm{ai}}\vert\phi_{n^{\prime}},\psi_{n^{\prime}}\rangle$ appears with $k=1,2$, 
\begin{align}
\langle\psi_{n}\vert\psi_{n^{\prime}}\rangle^{N-1}\mapsto \frac{\langle\psi_{n}\vert\psi_{n^{\prime}}\rangle^{N-1}}{\langle\psi_{n}\vert\psi_{n}\rangle^{N-1}}
\end{align}
for the term in which $\langle\phi_{n}\vert \bar{V}_{\mathrm{ai}}\vert\phi_{n^{\prime}}\rangle$ appears, and
\begin{align}
\langle\vert\psi_{n}\vert^{2}\rangle\mapsto \frac{\langle\vert\psi_{n}\vert^{2}\rangle}{\langle\psi_{n}\vert\psi_{n}\rangle}
\end{align}
for the terms in which $\overline{g}$ appears (this also applies for $H_{gp}^{n}[\psi_{n}]$). Note that for states normalized to unity the equations~(\ref{eq:eq19},\ref{eq:eq20}) are unchanged. Instead the appearance of the norm $\langle\psi_{n}\vert\psi_{n}\rangle$ is essential  in the TW, where the generated stochastic fields have a norm of the order $O(N)$. 


\subsection{Numerical procedure}

We have implemented numerically the equations~(\ref{eq:eq20}) in the following way: First, we have computed the wave function of the condensate by using the imaginary time propagation algorithm for the GPE~(\ref{eq:GPE}) by using as trial functions the single-particle ground and first excited states of the double-well potential~\cite{DalfovoPRA96}. These states are degenerate when the separation $2q$ between the wells is sufficiently large. The distance $q$ has been chosen in such a way that the atom-ion interaction is essentially negligible at $2q_0$, and therefore~(\ref{eq:eq20}) reduce to~(\ref{eq:GPE}). In this way, the left and right condensate wave functions are obtained by linear combination of the symmetric and antisymmetric solutions of the imaginary time propagation. For the latter we used a time step $\Delta t_{\mathrm{im}} = 10^{-4} \hbar/\bar{E}^*$. 
Once the initial left condensate wave function has been determined, we have expanded the state onto the eigenstates $\ket{m}$ of the double-well potential (i.e., $H_0$) at $q=3.5\bar{R}^{*}$ as
\begin{equation}
\label{eq:eq21}
\vert\psi_{n}\rangle=\sum_{m=0}^{n_{a}} B_{m}^{n}(t)\vert m\rangle.
\end{equation}
Here, we have truncated the Hilbert space in such a way that we safely consider the first $n_a+1$ eigenstates only. Thus, by inserting~(\ref{eq:eq21}) into~(\ref{eq:eq19}) we obtain:
\begin{equation}\label{eq:eq22}
\begin{aligned}
i\dot{C_{n}}=&N \sum_{n^{\prime}=0}^{n_{i}}C_{n^{\prime}}\left[\sum_{m,m^{\prime}=0}^{n_a}(B_{m}^{n})^*
B_{m^{\prime}}^{n^{\prime}}\langle m, n\vert \bar{V}_{\mathrm{ai}}\vert n^{\prime}, m^{\prime}\rangle\right. \\
&\left.\times\left(\sum_{m=0}^{n_{a}} (B_{m}^{n})^* B_{m}^{n^{\prime}}\right)^{(N-1)}\right].
\end{aligned}
\end{equation}
Finally, by substituting~(\ref{eq:eq21}) into~(\ref{eq:eq20}) and by multiplying $\langle m\vert$ from the left-hand side, we obtain the following final equations of motion of the new expansion coefficients: 
\begin{widetext}
\begin{eqnarray}\label{eq:eq23}
i\dot{B_{m}^{n}}=&& \left[E_{0}^{a}(m) + \dfrac{\overline{g}}{2}(N-1)B_{m}^{n}  + \dfrac{\bar{E}_{n}}{N}\right]B_{m}^{n} -\sum_{m^{\prime}}B_{m^{\prime}}^{n}\langle m\vert \bar{V}_{\mathrm{ext}}^{q=cons} - \bar{V}_{\mathrm{ext}}^{q(t)}\vert m^{\prime}\rangle\nonumber\\
&& +(N-1)\dfrac{C_{n}^{*}}{\vert C_{n}\vert^{2}}\left[\sum_{n^{\prime}=0}^{n_{i}} C_{n^{\prime}}\left(\sum_{m,m^{\prime}=0}^{n_a}(B_{m}^{n})^*
B_{m^{\prime}}^{n^{\prime}}\langle m, n\vert \bar{V}_{\mathrm{ai}}\vert n^{\prime}, m^{\prime}\rangle\right) 
\left(\sum_{m=0}^{n_{a}} (B_{m}^{n})^* B_{m}^{n^{\prime}}\right)^{(N-2)} B_{m}^{n^{\prime}}\right] \nonumber\\
&& +\dfrac{C_{n}^{*}}{\vert C_{n}\vert^{2}}\left[\sum_{n^{\prime}=0}^{n_{i}} C_{n^{\prime}}\left(\sum_{m^{\prime}}
B_{m^{\prime}}^{n^{\prime}}\langle m, n\vert \bar{V}_{\mathrm{ai}}\vert n^{\prime}, m^{\prime}\rangle\right)\left(\sum_{m=0}^{n_{a}} (B_{m}^{n})^* B_{m}^{n^{\prime}}\right)^{(N-1)}\right] \nonumber\\
&&-N\dfrac{C_{n}^{*}}{\vert C_{n}\vert^{2}}\left[\sum_{n^{\prime}=0}^{n_{i}} C_{n^{\prime}}\left(\sum_{m,m^{\prime}=0}^{n_a}(B_{m}^{n})^*
B_{m^{\prime}}^{n^{\prime}}\langle m, n\vert \bar{V}_{\mathrm{ai}}\vert n^{\prime}, m^{\prime}\rangle\right)\left(\sum_{m=0}^{N_{a}} (B_{m}^{n})^* B_{m}^{n^{\prime}}\right)^{(N-1)}\right] B_{m}^{n}.
\end{eqnarray}
\end{widetext}
Here $H_{0}\lvert m\rangle=E_{0}^{a}(m)\lvert m\rangle$, $\vert n\rangle\equiv\vert \phi_{n}\rangle$, and $V_{\mathrm{ext}}^{q=cons}$ is the added and subtracted double-well potential at a constant value $q=3.5\bar{R}^{*}$.
To solve the above outlined differential equations we have chosen a time step of at least $\Delta t_{\mathrm{re}} = 10^{-2} \hbar/\bar{E}^*$ 
and used standard routines for integrating coupled different equations in \small{MATLAB}.

\nocite{*}

\bibliography{mypaper}

\end{document}